\documentclass[prb,twocolumn,superscriptaddress]{revtex4}
\usepackage{amsmath, amsfonts,amssymb,color,graphicx,epsfig}

\begin{document}

\title{Photo--induced electron transfer in the strong coupling regime: Waveguide--plasmon polaritons}

\author{Peng Zeng}
\affiliation{School of Chemistry, The University of Melbourne, VIC 3010, Australia}
\author{Jasper Cadusch}
\affiliation{School of Physics, The University of Melbourne, VIC 3010, Australia}
\author{Debadi Chakraborty}
\affiliation{School of  Mathematics and Statistics, The University of Melbourne, VIC 3010, Australia}
\author{Trevor A. Smith}
\affiliation{School of Chemistry, The University of Melbourne, VIC 3010, Australia}
\author{ Ann Roberts}
\affiliation{School of Physics, The University of Melbourne, VIC 3010, Australia}
\author{John E. Sader}
\affiliation{School of  Mathematics and Statistics, The University of Melbourne, VIC 3010, Australia}
\author{Timothy J. Davis}
\affiliation{School of Physics, The University of Melbourne, VIC 3010, Australia}
\author{Daniel E. G\'omez}
\email{daniel.gomez@csiro.au}
\affiliation{CSIRO,  Private Bag 33, Clayton, VIC, 3168, Australia}
\affiliation{School of Physics, The University of Melbourne, VIC 3010, Australia}

\maketitle

\textbf{
Reversible exchange of photons between a material and an optical cavity can lead to  the formation of hybrid light--matter states where material properties such as the  work function\cite{Hutchison_AM2013a}, chemical reactivity\cite{Hutchison_ACIE2012a}, ultra--fast energy relaxation \cite{Salomon_ACIE2009a,Gomez_TJOPCB2012a} and  electrical conductivity\cite{Orgiu_NM2015a} of matter differ significantly to those of the same material in the absence of  strong interactions with the electromagnetic fields. 
Here we show  that strong light--matter coupling between confined photons on a semiconductor waveguide and localised plasmon resonances on metal nanowires modifies the efficiency of the photo--induced  charge--transfer rate of plasmonic derived (hot) electrons into accepting states in the semiconductor material.
Ultra--fast spectroscopy measurements reveal a strong correlation between the amplitude of the transient signals, attributed to electrons residing in the semiconductor, and the hybridization of waveguide and plasmon excitations. 
}

\begin{figure*}[tbph!]
\centering
\includegraphics[width=0.8\linewidth]{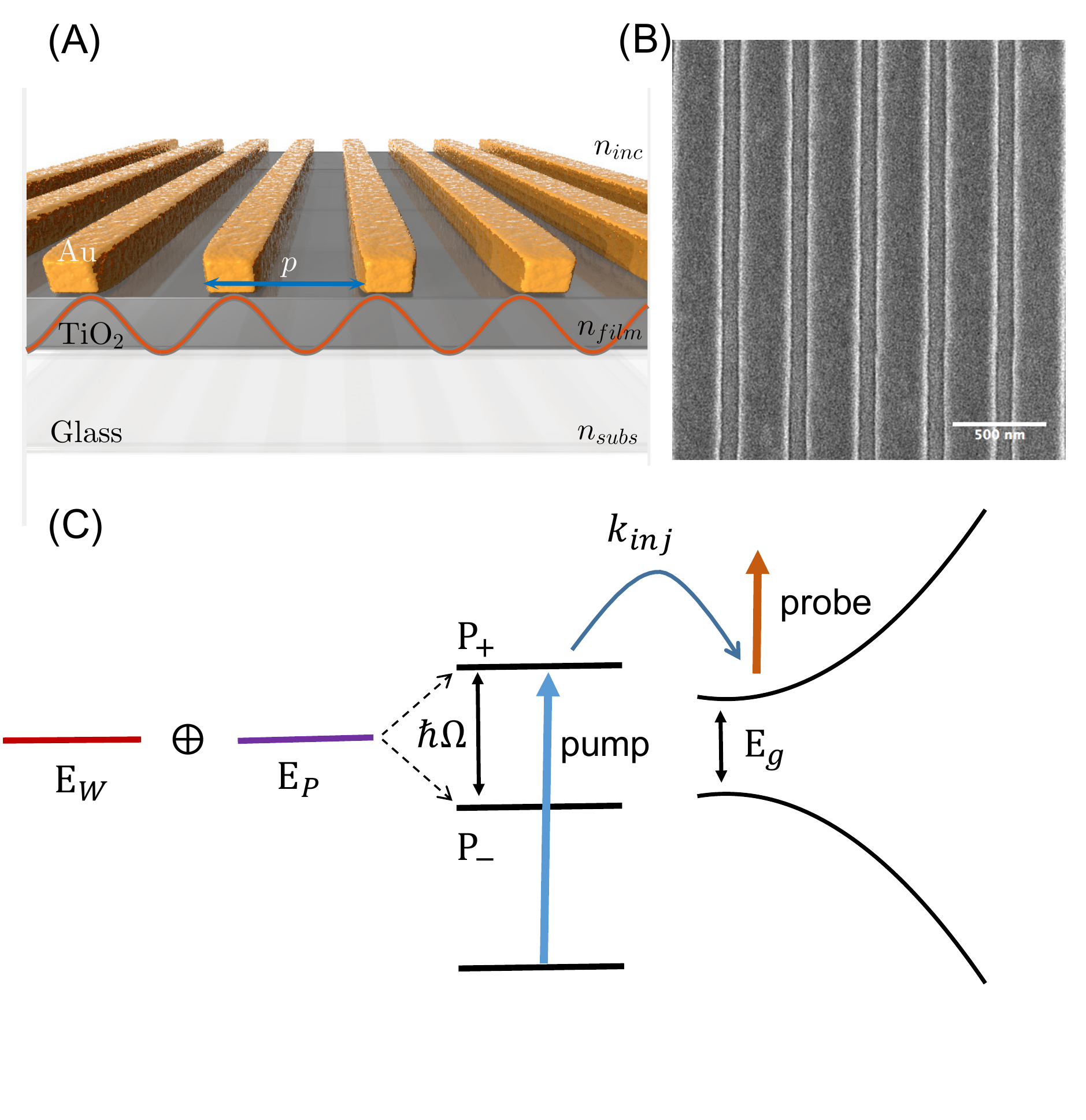}
\caption{\textbf{Strong waveguide--plasmon interactions} 
(A) Au gratings of 80 nm in width and 30 nm in height were deposited on top of 200 nm films of TiO$_2$. The film can support waveguide modes (red wavy lines) which when resonant with the localised plasmon resonances of the wires can lead to the formation of strongly coupled waveguide--plasmon polaritons. The excitation of these polaritons depends on the periodicity $p$ of the nanowire grating, for normally incident light. 
(B) SEM image of lithographically--made nanowires, the scale bar shown corresponds to 500 nm. 
(C) Diagram illustrating the formation of plasmon--waveguide hybrid states ($P_\pm$, split in energy by $\hbar\Omega$) from uncoupled waveguide and plasmon states characterised by having energies $E_W$ and $E_P$ respectively. 
The pump--probe experiments were carried out by pumping into $P_+$ states and probing the transient occupation of electrons in the conduction band of TiO$_2$ (represented by the curve vertically shifted by $E_g$ from the valence band).}
\label{fig:1}
\end{figure*}

The light--induced collective oscillation of charge carriers in metallic nanostructures -- plasmons-- results in enormous energy densities on the nanoscale which can greatly enhance a diverse range of photophysical processes\cite{Novotny_NP2011a}. 
Plasmon energy relaxation can result in the formation of energetic charge carriers which can generate localised heat through interactions with the host lattice \cite{Hartland_CR2011a} or charge--separated states with sufficient electrochemical potential to drive chemical reactions\cite{Zhang_ROPIP2013a,Clavero_NP2014a,Kale_AC2014a,Moskovits_NN2015a,Brongersma_NN2015a}.

Much effort has been focused on developing strategies to increase the yield of charge--carrier extraction from plasmonic structures.
This  yield  depends on the branching ratio between radiative and non--radiative plasmon decay.
The former  is partially controlled by the intrinsic geometry  of the metallic nanostructure \cite{Gomez_NL2013a} and  the density of photonic states around it.
Plasmons are predominantly radiatively--damped oscillations \cite{Sonnichsen_PRL2002a}.

On structures that can support both localised plasmon resonances and waveguide modes, it is possible to excite strongly--coupled admixtures of these two states resulting in \textit{waveguide--plasmon polaritons} (WPPs)\cite{Christ_PRL2003a,Christ_PRB2004a}.
These polaritons originate from the interaction of the evanescent fields from the  waveguided modes  and the (also  evanescent) fields from the particle plasmons.
These waveguide--plasmon polaritons are characterised by exhibiting sharp spectral linewidths that arise from the relatively long lifetimes of the photonic character of these polaritons, in addition to having highly localised near--fields around the metal structures.
The strong waveguide--like character of the waveguide--plasmon polaritons imply that in these systems there is a strong suppression of  radiative damping \cite{Zentgraf_PRL2004a}.

Ng {\it et al} \cite{Ng_AOM2015a} have shown a correlation between the periodicity of Al nanowire gratings supported on thin films of TiO$_2$ and the rate of photo--induced decomposition of methyl orange in solution, a process they postulated to be initiated by hot charge carrier separation at the metal--semiconductor Schottky contact. 
This process was argued to be enhanced when the metal nanowire grating exhibited strong coupling to waveguide modes in the supporting TiO$_2$ films.
Here, we show that strong coupling of particle plasmons and waveguide modes alters the injection of plasmonic hot--charge carriers into the supporting semiconductor waveguide.

\textbf{Steady state measurements: Existence of waveguide plasmon polaritons}.
\begin{figure*}[tbph!]
\centering
\includegraphics[width=0.9\linewidth]{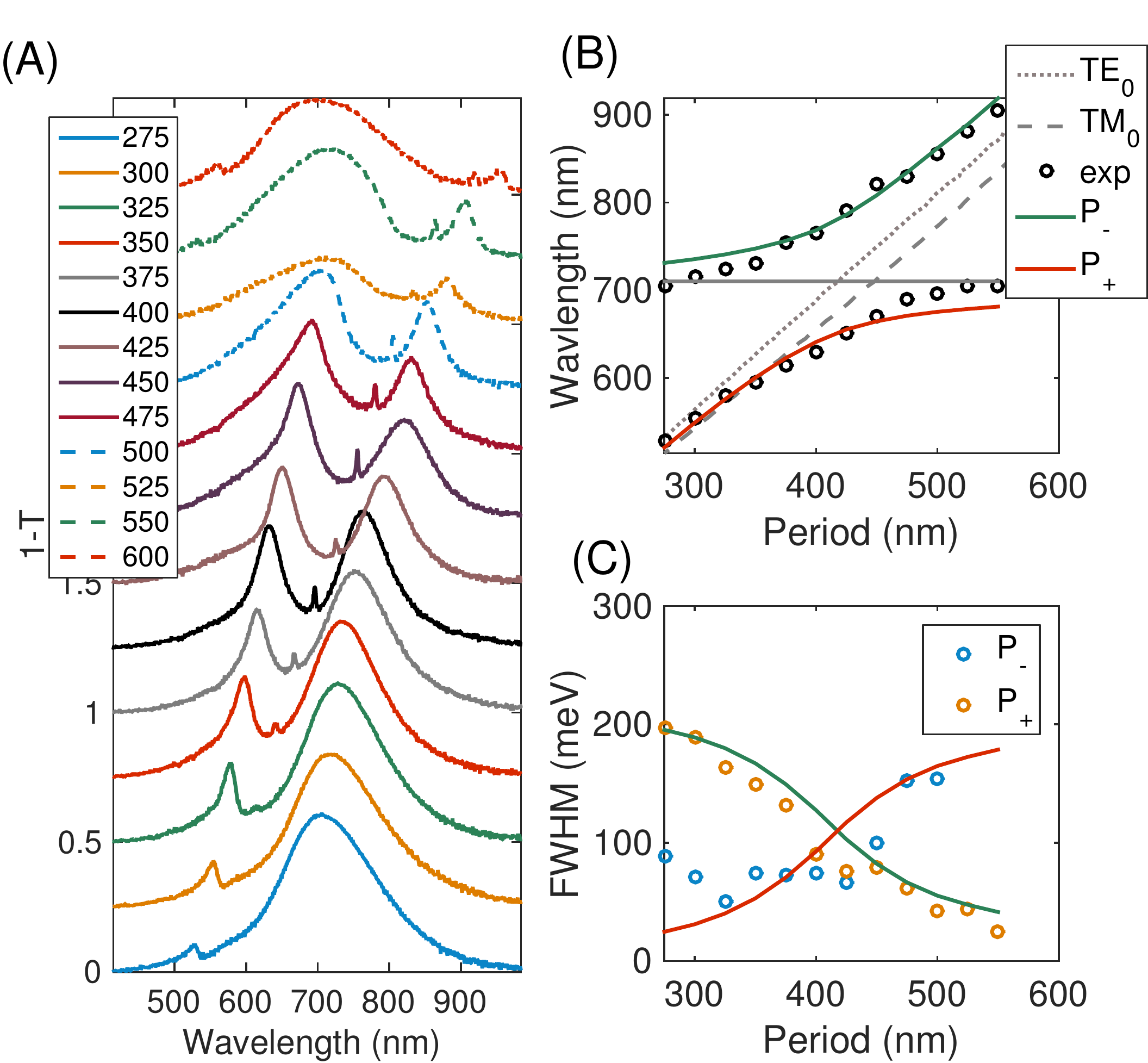}
\caption{\textbf{Steady state spectra and energy dispersion curve}. 
(A) Extinction data (1- Transmission) shown for several garting periods as indicated in the legend (in nm). There is a clear avoided crossing as the period of the nanowire grating increases. This indicates the formation of waveguide--plasmon polaritons. 
(B) Energy \textit{vs.} nanowire grating period curve. The dots were obtained from the maxima of the data shown in (A). The grey lines correspond to the uncoupled LSPR (continuous horizontal line) and the lowest--order TE and TM modes of the waveguide, calculated using Maxwell equations. The solid (coloured) lines correspond to the  eigen--energies of the polariton states  ($P_-$ and $P_+$) calculated using the coupled--oscillator model with $\hbar\Omega = 300$ meV. 
(C) Spectral linewidths. The dots show measured spectral linewidths for the upper and lower (in energy) branches whereas the continuous lines show the imaginary part of the  eigen--energies calculated ($P_+$ and $P_-$) using the coupled--oscillator model. }
\label{fig:2}
\end{figure*}
As shown in the diagram of figure \ref{fig:1}, the samples consist of Au  nanowire  gratings of variable period, deposited on top of a thin film of TiO$_2$.
Figure 1b shows a scanning electron microscope image of one such sample.
The measured extinction (1-Transmission) spectra of these samples, shown in figure \ref{fig:2}(A), were measured under normal incidence using light polarised perpendicular to the long axis of the nanowires.

Starting with a period of 250 nm, the measured spectrum  exhibits one peak due to the localised surface plasmon resonance (LSPR) of the metal nanostructure.
This assignment is supported by rigorous coupled wave analysis of these periodic structures \cite{Ng_AOM2015a} (see supplementary Information \ref{sec:RCWA}).
The grating can impart  momentum to the incident light and couple it to waveguide modes supported by the TiO$_2$ thin film.
As the grating period is increased, the waveguide mode can be brought into resonance with the LSPR mode.
The \textit{strong} interaction of the evanescent electromagnetic field of the $\text{TM}_0$ waveguide mode with  the (also evanescent) localised surface plasmon resonance of the nanowires results in spectral doublets that display an avoided crossing, as shown  in the progression of the spectra with increasing grating period  in figure \ref{fig:2}(A).
These doublets are characteristic of waveguide--plasmon polaritons: quasi--particles of the coupled system which can be described as a coherent superposition of 
plasmon and waveguide modes \cite{Christ_PRL2003a,Christ_PRB2004a}.
The physics of this interaction can be captured with a model of coupled harmonic oscillators \cite{Christ_PRL2003a, Christ_PRB2004a} which leads to the following (non--Hermitian) Hamiltonian (to be more accurate, the Hamiltonian should also include an additional entry due to the coupling and excitation of TE modes, see \cite{Christ_PRL2003a}):
\begin{equation}
\begin{pmatrix}
E_p-i\hbar\gamma_p & \hbar\Omega/2 \\
\hbar\Omega^*/2 & E_w-i\hbar\gamma_w
\end{pmatrix},
\label{eq:1}
\end{equation}
where $E_p$ ($E_w$) is the resonance energy of the LSPR (waveguide mode), $\gamma_p$ ($\gamma_w$) the damping frequency of the plasmon resonance (waveguide) and $\hbar\Omega$ is a potential describing the coupling strength between both oscillators, which dictates the energy splitting observed at resonance (when $E_p = E_w$, the situation depicted in figure 1(C)).
The eigen--energies resulting from a diagonalisation of this Hamiltonian with $\hbar\Omega$ = 300 meV are given by the continuous coloured lines shown in figure 2(B) which   satisfactorily reproduce the experimental data. 
This model predicts the formation of two polariton states: $P_-$ and $P_+$ as indicated in the diagram of figure 1(C), which are superpositions of plasmon--like ($|sp\rangle$) and waveguide--like $|w\rangle$ states (i.e. $|P_+\rangle = c_p^+|sp\rangle + c_W^+|w\rangle$, more details in SI \ref{sec:S_oscillators}).
These states have properties that differ from those of the uncoupled sub--systems, including their spectral linewidths, shown in figure 2(C).

According to this simple model, incident light energy can  be reversibly exchanged between the plasmon-- and waveguide--like character of the polaritons \cite{Vasa_NP2013a}, which results in longer dephasing times for the plasmon--like excitations \cite{Zentgraf_PRL2004a} and in principle could also lead to  the observation of  polariton beats \cite{Zentgraf_PRL2004a,Vasa_NP2013a}.
In our samples, the energies of the polariton states $P_+$ are above the Au--TiO$_2$ Schottky barrier ($\sim$ 1 eV \cite{Zhang_CR2012a}) and one possible relaxation mechanism for these polariton states is that of electron transfer from the Au nanowires into the conduction band of TiO$_2$ \cite{Clavero_NP2014a, Moskovits_NN2015a,Brongersma_NN2015a}.

Electron transfer from plasmonic nanostructures into supporting semiconductor materials is thought to occur following non--radiative (Landau) damping of plasmons \cite{Brongersma_NN2015a}. 
Landau damping can take place on a time scale between 1 - 100 fs following light excitation.
It occurs in competition with radiative damping (re--emission of a photon) and is argued to result in the formation of a non--equilibrium electron--hole pair in the metal (hot charge carriers).
In the subsequent 100 fs to 1 ps the non--equilibrium charge carriers relax via electron--electron scattering processes and at this stage the energy can be transferred to the surrounding medium (or lattice) as heat \cite{Hartland_CR2011a} or can be imparted to a single electron as kinetic energy resulting in transport into the adjacent semiconductor support.

Injected electrons  in the conduction band of TiO$_2$ result in  reduced TiO$_2$ species which are  known to absorb both visible and near--IR light \cite{Tachikawa_TJOPCB2006a} enabling thus the detection of these charge transfer events by means of visible pump near--IR probe spectroscopy as described next.

\textbf{Ultra--fast pump--probe measurements.}
\begin{figure}[tbph!]
\centering
\includegraphics[width=0.9\linewidth]{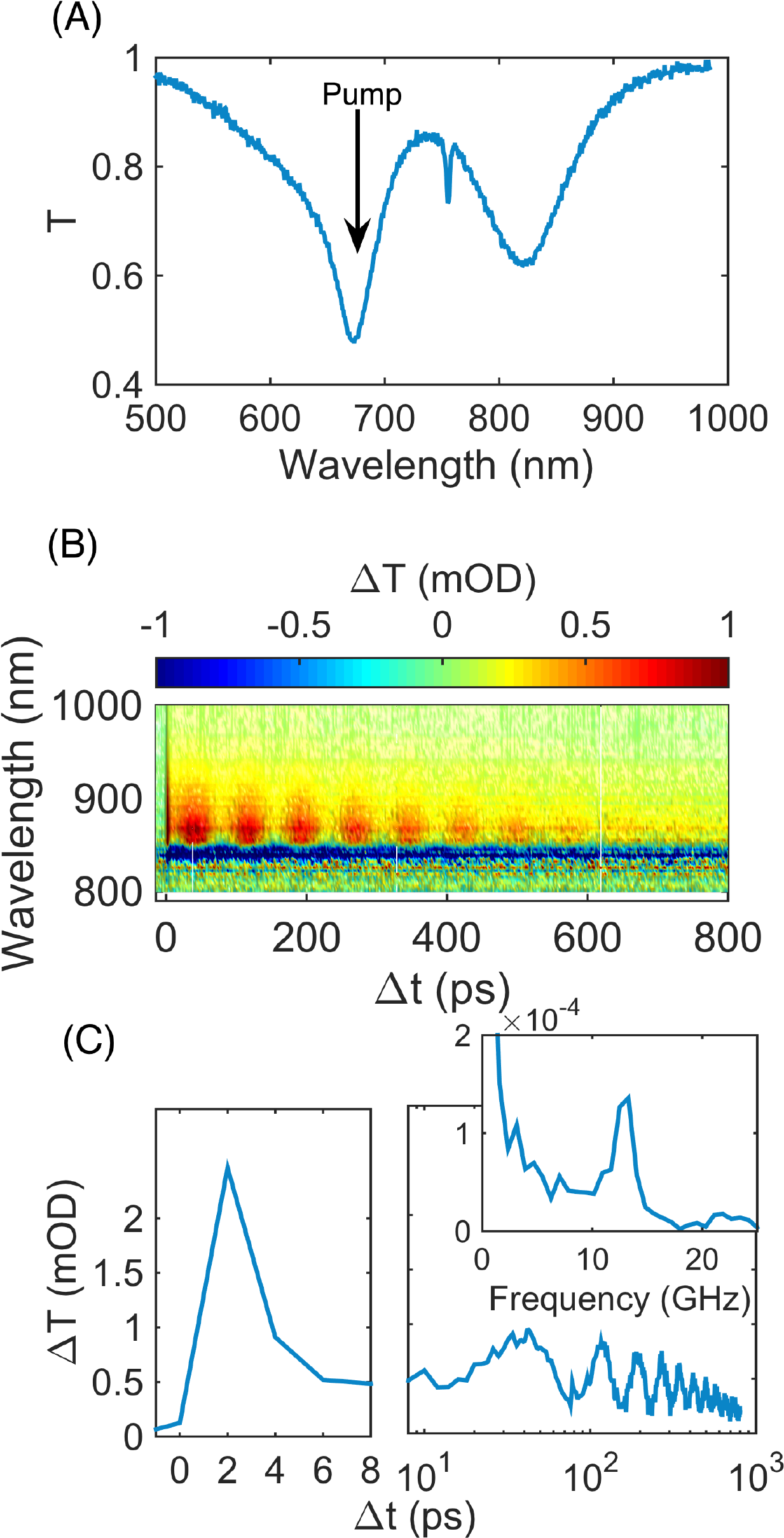}
\caption{\textbf{Femtosecond Pump -- Probe spectroscopy} (A) Steady state extinction spectrum of a sample with a period of 425 nm. The arrow indicates the spectral location of the pump pulses. (B)  Difference transmission spectra as a function of probe wavelength and pump--probe delay time (C) Kinetic trace measured at a probe wavelength of 868 nm. The inset shows the Fourier Transform of the oscillatory part of the kinetic trace, clearly revealing a component located at $\approx$ 13.3GHz  }
\label{fig:3}
\end{figure}
In order to study the kinetics of the relaxation of waveguide plasmon polaritons, we excited the samples using ultra--short laser pulses spectrally tuned to the higher--energy extinction bands of the WPPs (or $P_+$ states, blue arrow in figure1(C)) and monitored changes in the transmission spectrum of broad--band light pulses with a controlled time delay with respect to the arrival of the pump pulses (red arrow, figure 1(C)).
The pump pulses were lightly focused at normal incidence to the nanowire gratings and were also polarised perpendicular to the long axis of the nanowires. 
The incident power was adjusted to be as low as possible (whilst attaining sufficient signal--to--noise) to ensure that the photo--induced processes were linear with the excitation power (indeed the magnitude of the signals increased linearly with excitation power, see SI \ref{sec:sTA}).

Figure \ref{fig:3} shows the results of these measurements for the sample with a period of 425 nm (close to the plasmon--waveguide resonance point, see fig. \ref{fig:2}(B)) (the spectra for other grating periods are shown in the SI, along with those obtained for the bare TiO$_2$ substrate ).
The data of figure \ref{fig:3}(B) clearly shows that there are transient spectral changes in the region between 800 -- 900 nm.
Spectrally, the measured signals are dominated by  both a positive and negative contributions around the position of the lower polariton band.
Similar observations have been made for the case of strongly coupled plasmon--exciton systems \cite{Schwartz_C2013a,Salomon_ACIE2009a, Gomez_TJOPCB2012a} which have been interpreted as originating from transient changes in the dielectric constants of the materials.
In the time domain, figure \ref{fig:3}(C) shows that the detected signals  consist of a fast rise and  rapid decay ($\Delta t <$ 5 ps) which is followed by a slower decay ($\sim$ 100s ps) that exhibits oscillations for some spectral regions (see figure \ref{fig:3}(B)). 
A Fourier transform of these oscillations reveals the presence of a harmonic at $\sim$ 13.3 GHz which originates from   the coherent excitation of (mechanical) breathing vibrational modes of the nanowires (see SI  \ref{sec:SI_mech}).

The transient occupation of electrons in the conduction band of TiO$_2$ results in a broadband transient signal that takes place between 600 nm and all the way up to 3000 nm \cite{Bian_JOTACS2014a,Furube_JOTACS2007a}.
Therefore, the measured changes in transmission at wavelengths $>$ than 900 nm report on the electron population in the conduction band of TiO$_2$.
Evidence that these signals originate from these electron injection events is found in figure 4(A) where we show transient spectroscopy results for nanowire gratings deposited on an inert (i.e. no electron accepting states) glass substrate, in addition to a nanowire grating deposited on a TiO$_2$ substrate using a Ti adhesion layer, which has been shown to result in an Ohmic contact \cite{Zheng_NC2015a}.
As shown on the schematic at the top of figure 4, the nature of the different contacts and substrates is expected to lead to marked differences in the transient spectra.
For a glass substrate, electron injection from Au nanowires is unlikely due to the unfavourable energetics.
For an Ohmic contact, whilst the electron injection is possible, the lifetime of the charge separated states will be significantly smaller in comparison to those of the Schottky contacts, where due to band--bending, an energetic barrier (Schottky barrier, with height $\phi_{SB}$ in figure 4) prevents the injected electron from directly recombining with the hole left in the metal.
These differences translate into the observed changes in the amplitude of the transient signals shown in figure 4(A).
Therefore, at long wavelengths, the measured transient changes in transmission report on the transient population of electrons in the TiO$_2$ that originate from non--radiative decay of the plasmon--like excitations of the $P_+$ state.
Figure 4(B) shows kinetic traces of these signals for several nanowire grating periods.

A simple kinetic model (developed in section \ref{sec:kinetic_model}, see equation \ref{eq:SNa}) shows that the amplitude of the signals shown in Figure 4(B) are proportional to the quantum yield for electron injection and the plasmon--like character of the $P_+$ state, dictated in turn by the coefficient $c_p^+$:
the plasmon component of an initially populated $P_+$ excited state can depopulate by electron transfer from the metal nanowire into accepting states in TiO$_2$.
Figure 4(C) shows the amplitude of these signals (measured at $\Delta t$=0 in figure 4(B)) vs grating period, which clearly shows that this amplitude changes with the period of the grating.
We argue that the evolution of the amplitude of the transient signals is due to an interplay between a reduced radiative damping due to plasmon--waveguide hybridization and the plasmon character of the resulting hybrid state, as we discuss next.

\begin{figure}[tbph!]
\centering
\includegraphics[width=0.9\linewidth]{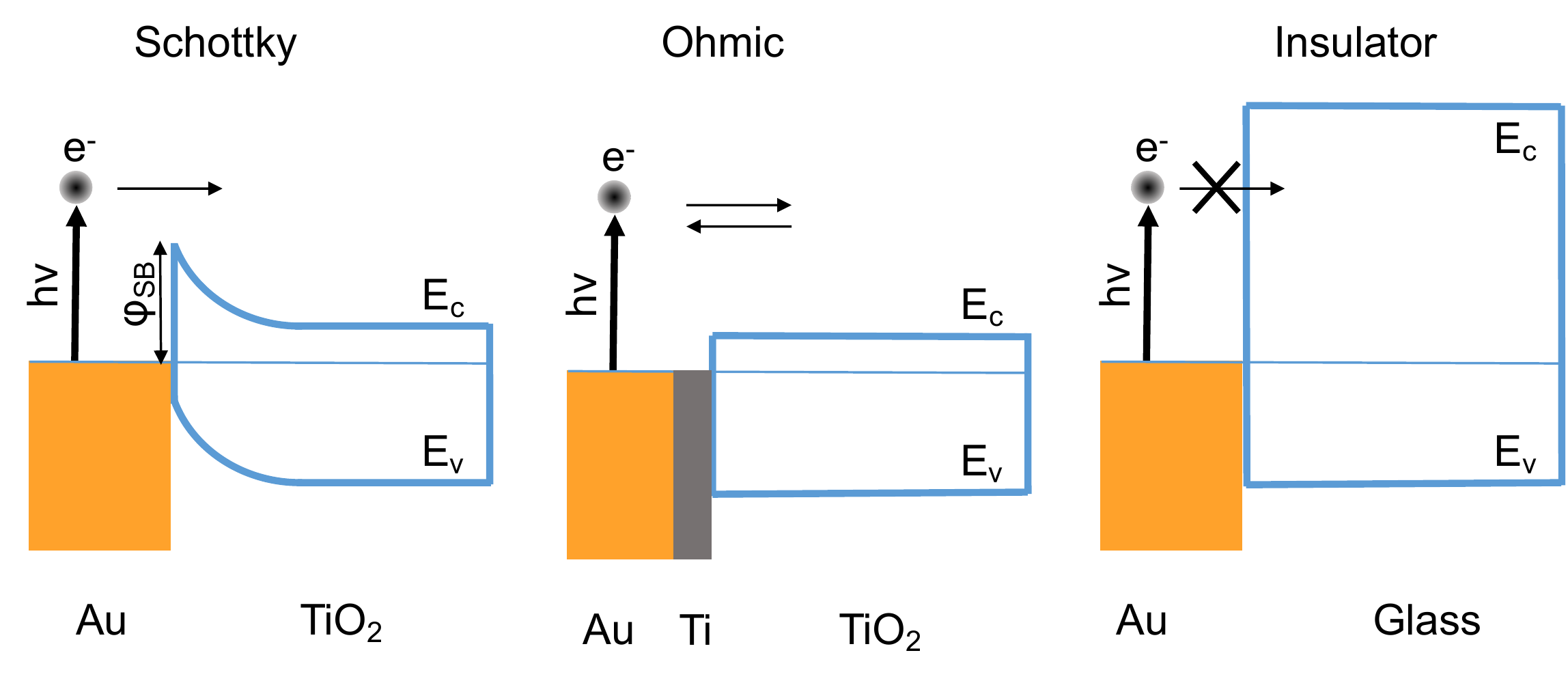}\\
\includegraphics[width=0.9\linewidth]{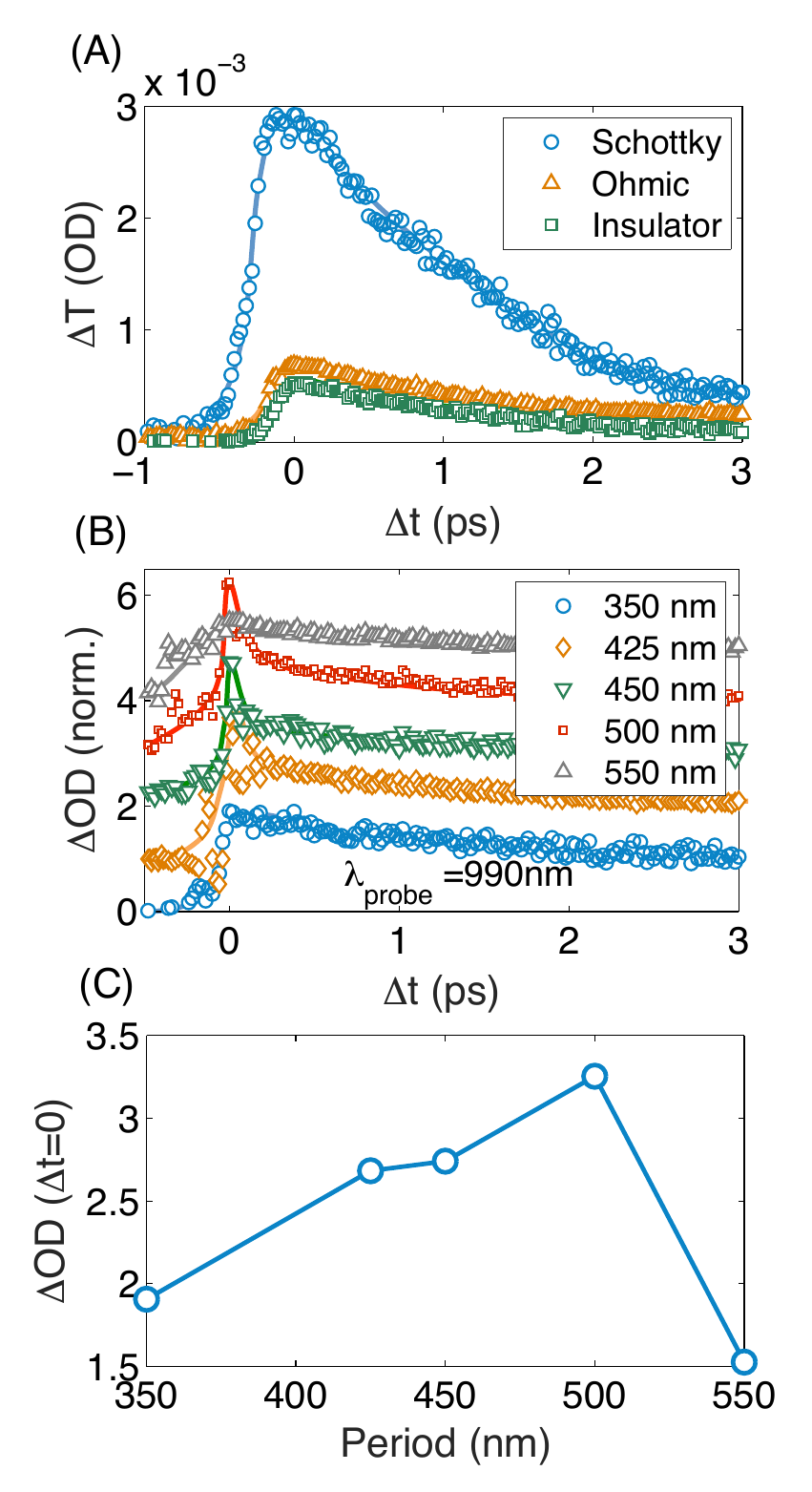}
\caption{\textbf{Electron injection} 
Diagrams representing the differences in the electrical nature of the contacts of (A): On an insulating substrate, electron injection is not readily possible. On Ohmic contacts electron back transfer from semiconductor to metal can take place limiting the lifetime of the charge--separated state. On Schottky contacts the barrier slows down this back reaction. 
(A) Transient spectra for Au nanowire gratings deposited directly on thin films of TiO$_2$ forming a Schottky contact, deposited with a 2nm Ti adhesion layer, leading to an Ohmic contact  \cite{Zheng_NC2015a} and on a glass (insulator) substrate. 
(B) Difference transmission spectra measured at a common probe wavelength removed from polariton branches. 
(C) Maximum amplitude of the data shown in (B) plotted as a function of grating period.}
\label{fig:4}
\end{figure}
The quantum yield for electron injection $\phi$ is given by $\phi =k_{inj}/(k_{inj}+k_{rad}+k_{nr})$, where $k_{inj}$ is the rate of electron injection, $k_{rad}$ the rate of radiative damping of the plasmons and $k_{nr}$ is the rate of all other non--radiative decay channels.
$k_{inj}$ depends on the overlap of the wavefunctions of the electron donating and accepting states, the density of accepting states, the reorganisation energy and the free energy associated with the electron transfer. 
Except for the last parameter, these are all largely controlled by the nature of the Au/TiO$_2$ interface, which remains unchanged as the period is varied  in our experiments.
In the absence of electron transfer, the total dephasing rate of a plasmon resonance is given by $k_{rad}+k_{nr}$, a rate that is substantially modified when the plasmons hybridise with waveguide modes as shown experimentally by Zentgraf {\it et al}\cite{Zentgraf_PRL2004a}. 
The plasmon radiative decay rate $k_{rad}$ can be approximated as\cite{Stanley_PRB1996a}:
\begin{equation}
k_{rad} \propto |\langle sp|H_{rad}|\text{final state}\rangle|^2
\end{equation}
where $H_{rad}$ is the Hamiltonian describing the radiative scattering of plasmons and $|sp\rangle$ denotes an initial state of the metallic nanostructure supporting a surface plasmon oscillation.
For a strongly coupled waveguide--plasmon system, the states are given by coherent superpositions (i.e. $|P\rangle=c_w|w\rangle + c_p|sp\rangle$. More details in section \ref{sec:S_oscillators}) and therefore the radiative decay rate would be modified to a value $\bar{k}_{rad}$ given instead by:
\begin{equation}
\begin{split} 
\bar{k}_{rad} &\propto \bigg|\left( \langle sp|c_p^* \pm \langle w|c_w^* \right)|H_{rad}|\text{final state}\rangle\bigg|^2\\
&\propto |c_p|^2 k_{rad}
\end{split}
\label{eq:3}
\end{equation}
which, since $|c_p|^2\le 1$, implies that waveguide--plasmon strong coupling reduces the rate of  radiative damping and consequently increases the electron injection quantum yield $\phi$ (for an alternate explanation, see SI \ref{sec:S_rad}).
The magnitude of the measured signal of figure 4(B) is proportional to the product of the  quantum yield of electron injection {\it and} the plasmon character $|c_p^+|^2$ of the $P_+$ excited state. 
The later increases with increasing grating period (see figure 2(B) and SI figure \ref{fig:SFig2}), but the former, according to equation \eqref{eq:3} and the definition of the quantum yield, decreases with grating period. 
The net result is that there is an optimum point of plasmon--waveguide hybridisation that leads to increased electron injection, as shown by the data of figure 4(C).

Our work demonstrates that modifications of the electromagnetic environment of plasmonic systems, and in particular  strong coupling of these excitations with electromagnetic fields leads to a modification of the rate (efficiency) of plasmonic hot--charge carrier injection.
Strategies to increase the yield of  metal--to--semiconductor charge transfer are important for the application of this concept in solar energy harvesting with plasmonic nanostructures\cite{Moskovits_NN2015a}: both in photovoltaics \cite{Mubeen_AN2014a} and photocatalysis \cite{Mubeen_NN2013a,Linic_NM2011a,Zhang_ROPIP2013a} and also for the development of novel concepts in opto--electronics\cite{Abb_NL2011a}.


\section{Methods}

\textit{Sample preparation}
TiO$_2$ films were deposited on glass substrates by means of  electron beam evaporation at $\sim$ 1.5 \AA{}/sec rate to a nominal mass thickness of 200 nm.
The films were annealed under an oxygen atmosphere at 450 $^\circ$C after deposition.
Nanowire gratings were fabricated by means of electron beam lithography, using a Vistec EBPG 5000 plus ES (100 keV, 3 nA). 
To this end, the TiO$_2$--coated glass substrates were covered with a  double--layer positive resist consisting of 50 nm of poly(methyl methacrylate) (Micro-Chem, 950k A2) on top of 110 nm of methyl methacrylate (Micro-Chem, MMA(8.5) MAA EL6).
A 20 nm layer of Cr was deposited on top of the resist in order to provide a charge dissipation layer during the electron beam lithography step. 
After exposure, the Cr layer was chemically etched and 
the patterns on the positive resist layer  were developed with a 1:3 methyl isobutyl ketone/ isopropanol solution for 60 s, rinsed by ultrasonication  with isopropanol, and dried with a nitrogen gun. 
2 nm of Ti followed by 30 nm of Au were deposited by electron beam evaporation at $\sim$ 0.5 \AA{}/sec rate. 
The final step in the nanofabrication consisted on a lift off step with acetone.
The structures were characterised by scanning electron microscopy (FEI, Nova NanoSEM 430) and had lateral extensions of 500 $\mu$m x 500$\mu$m.

\textit{Steady--state extinction measurements}
Light from a halogen source was polarised with a  polariser and was incident normal to the surface of the samples.
The transmitted light was collected using a low magnification objective lens using an inverted microscope and  was sent to the entrance port of an imaging spectrometer (Andor Shamrock SR-303i-A) equipped with a CCD (Andor iDus DU920P-BR-DD).
Transmission spectra were measured relative to bare TiO$_2$ areas in the sample.

\textit{Femtosecond Transient Absorption Spectroscopy}
A high--repetition--rate regenerative amplifier (Coherent RegA9050, 800 nm, 1W and 96 kHz) was employed as the laser source. 
The output of the RegA amplifier was re--compressed (Coherent EC9150) to 60 fs before being split to generate pump and probe pulses. 
A branch of the output pumped a tunable optical parametric amplifier (Coherent OPA9450) to produce pump pulses in visible range (480--750 nm).  
Near-IR white light probe pulses (800--1400 nm) were generated by focusing the rest of fundamental pulses (800 nm) into a 3 mm--thick YAG window (CASTECH).
A longpass filter was used to block the visible light in the probe pulses. 
Pump and probe pulses were focused on the sample with an off-axis parabolic mirror and overlapped at the sample with a pump spot size of 200 $\mu$m in diameter. The pump pulses were normally incident on the sample and polarised perpendicular to the long axis of the nanowires.
 
The absorption changes were measured by comparing adjacent probe pulses with and without pump pulses using a synchronised mechanical chopper in the path of the pump beam. 
The time--resolved transient absorption spectra were recorded using 
visible and NIR versions of the high-speed fiber-optic spectrometer (Ultrafast System, $\sim$9,200 and 7,100 spectra/sec, respectively). 
The delay time between pump and probe pulses was controlled using a motorized delay stage (Newport UTM-PP0.1, with step size of 0.66 fs and range of 800 ps). 
The temporal resolution of the whole setup was estimated at 200 fs (FWHM) by measuring the auto--correlation of fundamental pulses. 
A low noise level was achieved below 5x10$^{-5}$ OD, by taking  advantage of the high--repetition--rate laser and high signal averaging approach coupled with an acquisition process that rejected any outlying spectra. 
Photo-degradation and thermal effects were avoided by using low incident powers and their absence was ascertained by repeating the measurements several time ensuring identical outcomes. 



\begin{thebibliography}{10}

\bibitem{Hutchison_AM2013a}
Hutchison, J.~A., Liscio, A., Schwartz, T., Canaguier-Durand, A., Genet, C.,
  Palermo, V., Samor{\`\i}, P., and Ebbesen, T.~W.
\newblock ``Tuning the work-function via strong coupling,''.
\newblock {\em Advanced Materials}{ \bf 25}(17), 2481--2485 (2013).

\bibitem{Hutchison_ACIE2012a}
Hutchison, J.~A., Schwartz, T., Genet, C., Devaux, E., and Ebbesen, T.~W.
\newblock ``Modifying chemical landscapes by coupling to vacuum fields,''.
\newblock {\em Angewandte Chemie International Edition}{ \bf 51}(7), 1592--1596
  (2012).

\bibitem{Salomon_ACIE2009a}
Salomon, A., Genet, C., and Ebbesen, T.
\newblock ``Molecule--light complex: Dynamics of hybrid molecule--surface
  plasmon states,''.
\newblock {\em Angewandte Chemie International Edition}{ \bf 48}(46),
  8748--8751 (2009).

\bibitem{Gomez_TJOPCB2012a}
Gomez, D., Lo, S.~S., Davis, T.~J., and Hartland, G.~V.
\newblock ``Pico--second kinetics of strongly coupled excitons and surface
  plasmon polaritons,''.
\newblock {\em The Journal of Physical Chemistry B}{ \bf 117}, 4340--4346
  (2013).

\bibitem{Orgiu_NM2015a}
Orgiu, E., George, J., Hutchison, J.~A., Devaux, E., Dayen, J.~F., Doudin, B.,
  Stellacci, F., Genet, C., Schachenmayer, J., Genes, C., Pupillo, G., Samori,
  P., and Ebbesen, T.~W.
\newblock ``Conductivity in organic semiconductors hybridized with the vacuum
  field,''.
\newblock {\em Nat Mater}{ \bf 14}(11), 1123--1129 11  (2015).

\bibitem{Novotny_NP2011a}
Novotny, L. and van Hulst, N.
\newblock ``Antennas for light,''.
\newblock {\em Nat Photon}{ \bf 5}(2), 83--90 02  (2011).

\bibitem{Hartland_CR2011a}
Hartland, G.~V.
\newblock ``Optical studies of dynamics in noble metal nanostructures,''.
\newblock {\em Chemical Reviews}{ \bf 111}(6), 3858--3887 06  (2011).

\bibitem{Zhang_ROPIP2013a}
Zhang, X., Chen, Y.~L., Liu, R.-S., and Tsai, D.~P.
\newblock ``Plasmonic photocatalysis,''.
\newblock {\em Reports on Progress in Physics}{ \bf 76}(4), 046401 (2013).

\bibitem{Clavero_NP2014a}
Clavero, C.
\newblock ``Plasmon-induced hot-electron generation at nanoparticle/metal-oxide
  interfaces for photovoltaic and photocatalytic devices,''.
\newblock {\em Nat Photon}{ \bf 8}(2), 95--103 02  (2014).

\bibitem{Kale_AC2014a}
Kale, M.~J., Avanesian, T., and Christopher, P.
\newblock ``Direct photocatalysis by plasmonic nanostructures,''.
\newblock {\em ACS Catalysis}{ \bf 4}(1), 116--128 (2014).

\bibitem{Moskovits_NN2015a}
Moskovits, M.
\newblock ``The case for plasmon-derived hot carrier devices,''.
\newblock {\em Nat Nano}{ \bf 10}(1), 6--8 01  (2015).

\bibitem{Brongersma_NN2015a}
Brongersma, M.~L., Halas, N.~J., and Nordlander, P.
\newblock ``Plasmon-induced hot carrier science and technology,''.
\newblock {\em Nat Nano}{ \bf 10}(1), 25--34 01  (2015).

\bibitem{Gomez_NL2013a}
G\'omez, D.~E., Teo, Z.-Q., Altissimo, M., Davis, T., Earl, S., and Roberts, A.
\newblock ``The dark side of plasmonics,''.
\newblock {\em Nano Letters}{ \bf 13}, 3722--3728 (2013).


\bibitem{Sonnichsen_PRL2002a}
S\"onnichsen, C.,  Franzl, T.,  Wilk, T.,  von Plessen, G.,  Feldmann, J.,  Wilson, O., and Mulvaney, P.,
\newblock ``Drastic Reduction of Plasmon Damping in Gold Nanorods'' .
\newblock {\em Phys. Rev. Lett.}{\bf 88}, 077402 (2002)


\bibitem{Christ_PRL2003a}
Christ, A., Tikhodeev, S.~G., Gippius, N.~A., Kuhl, J., and Giessen, H.
\newblock ``Waveguide-plasmon polaritons: Strong coupling of photonic and
  electronic resonances in a metallic photonic crystal slab,''.
\newblock {\em Phys. Rev. Lett.}{ \bf 91}, 183901 Oct  (2003).

\bibitem{Christ_PRB2004a}
Christ, A., Zentgraf, T., Kuhl, J., Tikhodeev, S.~G., Gippius, N.~A., and
  Giessen, H.
\newblock ``Optical properties of planar metallic photonic crystal structures:
  Experiment and theory,''.
\newblock {\em Phys. Rev. B}{ \bf 70}, 125113 Sep  (2004).

\bibitem{Zentgraf_PRL2004a}
Zentgraf, T., Christ, A., Kuhl, J., and Giessen, H.
\newblock ``Tailoring the ultrafast dephasing of quasiparticles in metallic
  photonic crystals,''.
\newblock {\em Phys. Rev. Lett.}{ \bf 93}, 243901 Dec  (2004).

\bibitem{Ng_AOM2015a}
Ng, C., Dligatch, S., Amekura, H., Davis, T.~J., and G\'omez, D.~E.
\newblock ``Waveguide---plasmon---polariton enhanced photochemistry,''.
\newblock {\em Advanced Optical Materials}{ \bf } (2015).

\bibitem{Vasa_NP2013a}
Vasa, P., Wang, W., Pomraenke, R., Lammers, M., Maiuri, M., Manzoni, C.,
  Cerullo, G., and Lienau, C.
\newblock ``Real-time observation of ultrafast rabi oscillations between
  excitons and plasmons in metal nanostructures with j-aggregates,''.
\newblock {\em Nat Photon}{ \bf 7}(2), 128--132 02  (2013).

\bibitem{Zhang_CR2012a}
Zhang, Z. and Yates, J.~T.
\newblock ``Band bending in semiconductors: Chemical and physical consequences
  at surfaces and interfaces,''.
\newblock {\em Chemical Reviews}{ \bf 112}(10), 5520--5551 (2012).

\bibitem{Tachikawa_TJOPCB2006a}
Tachikawa, T., Tojo, S., Fujitsuka, M., Sekino, T., and Majima, T.
\newblock ``Photoinduced charge separation in titania nanotubes,''.
\newblock {\em The Journal of Physical Chemistry B}{ \bf 110}(29), 14055--14059
  (2006).

\bibitem{Schwartz_C2013a}
Schwartz, T., Hutchison, J.~A., L{\'e}onard, J., Genet, C., Haacke, S., and
  Ebbesen, T.~W.
\newblock ``Polariton dynamics under strong light--molecule coupling,''.
\newblock {\em ChemPhysChem}{ \bf 14}(1), 125--131 (2013).

\bibitem{Bian_JOTACS2014a}
Bian, Z., Tachikawa, T., Zhang, P., Fujitsuka, M., and Majima, T.
\newblock ``Au/Tio2 superstructure-based plasmonic photocatalysts exhibiting
  efficient charge separation and unprecedented activity,''.
\newblock {\em Journal of the American Chemical Society}{ \bf 136}(1), 458--465
  (2014).

\bibitem{Furube_JOTACS2007a}
Furube, A., Du, L., Hara, K., Katoh, R., and Tachiya, M.
\newblock ``Ultrafast plasmon-induced electron transfer from gold nanodots into
  {TiO$_2$} nanoparticles,''.
\newblock {\em Journal of the American Chemical Society}{ \bf 129}(48),
  14852--14853 (2007).

\bibitem{Zheng_NC2015a}
Zheng, B.~Y., Zhao, H., Manjavacas, A., McClain, M., Nordlander, P., and Halas,
  N.~J.
\newblock ``Distinguishing between plasmon-induced and photoexcited carriers in
  a device geometry,''.
\newblock {\em Nat Commun}{ \bf 6} 07  (2015).

\bibitem{Stanley_PRB1996a}
Stanley, R.~P., Houdr\'e, R., Weisbuch, C., Oesterle, U., and Ilegems, M.
\newblock ``Cavity-polariton photoluminescence in semiconductor microcavities:
  Experimental evidence,''.
\newblock {\em Phys. Rev. B}{ \bf 53}, 10995--11007 Apr  (1996).

\bibitem{Mubeen_AN2014a}
Mubeen, S., Lee, J., Lee, W.-r., Singh, N., Stucky, G.~D., and Moskovits, M.
\newblock ``On the plasmonic photovoltaic,''.
\newblock {\em ACS Nano}{ \bf 8}(6), 6066--6073 (2014).


\bibitem{Mubeen_NN2013a}
Mubeen, S., Lee, J., Singh, N., Kramer, S., Stucky, G.~D., and Moskovits, M.
\newblock ``An autonomous photosynthetic device in which all charge carriers
  derive from surface plasmons,''.
\newblock {\em Nat Nano}{ \bf 8}(4), 247--251 04  (2013).

\bibitem{Linic_NM2011a}
Linic, S., Christopher, P., and Ingram, D.~B.
\newblock ``Plasmonic-metal nanostructures for efficient conversion of solar to
  chemical energy,''.
\newblock {\em Nat Mater}{ \bf 10}(12), 911--921 12  (2011).

\bibitem{Abb_NL2011a}
Abb, M., Albella, P., Aizpurua, J., and Muskens, O.~L.
\newblock ``All-optical control of a single plasmonic nanoantenna--{ITO}
  hybrid,''.
\newblock {\em Nano Letters}{ \bf 11}, 2457 05  (2011).


\end{thebibliography}

\begin{thebibliography}
content...

\bibitem{Lalanne_JOMO1998a}
Lalanne, P. and Jurek, M.~P.
\newblock ``Computation of the near-field pattern with the coupled-wave method
  for transverse magnetic polarization,''.
\newblock {\em Journal of Modern Optics}{ \bf 45}(7), 1357--1374 (1998).

\bibitem{Lalanne_JOSAA1996a}
Lalanne, P. and Morris, G.~M.
\newblock ``Highly improved convergence of the coupled-wave method for tm
  polarization,''.
\newblock {\em J. Opt. Soc. Am. A}{ \bf 13}(4), 779--784 Apr  (1996).

\bibitem{Moharam_JOSAA1995a}
Moharam, M.~G., Gaylord, T.~K., Grann, E.~B., and Pommet, D.~A.
\newblock ``Formulation for stable and efficient implementation of the rigorous
  coupled-wave analysis of binary gratings,''.
\newblock {\em J. Opt. Soc. Am. A}{ \bf 12}(5), 1068--1076 May  (1995).

\bibitem{Rakic_AO1998a}
Raki\'{c}, A.~D., Djuri\v{s}i\'{c}, A.~B., Elazar, J.~M., and Majewski, M.~L.
\newblock ``Optical properties of metallic films for vertical-cavity
  optoelectronic devices,''.
\newblock {\em Appl. Opt.}{ \bf 37}(22), 5271--5283 Aug  (1998).

\bibitem{Wang_AN2014a}
Wang, W., Vasa, P., Pomraenke, R., Vogelgesang, R., De~Sio, A., Sommer, E.,
  Maiuri, M., Manzoni, C., Cerullo, G., and Lienau, C.
\newblock ``Interplay between strong coupling and radiative damping of excitons
  and surface plasmon polaritons in hybrid nanostructures,''.
\newblock {\em ACS Nano}{ \bf 8}(1), 1056--1064 01  (2014).

\bibitem{Tian_JOPAPAC2008a}
Tian, Y., Wang, X., Zhang, D., Shi, X., and Wang, S.
\newblock ``Effects of electron donors on the performance of plasmon-induced
  photovoltaic cell,''.
\newblock {\em Journal of Photochemistry and Photobiology A: Chemistry}{ \bf
  199}(2--3), 224 -- 229 (2008).

\end{thebibliography}


\section{Acknowledgements}
We would like to thank Gregory Hartland and James Hutchisnon for fruitful discussions on this work.
This work was performed in part at the Melbourne Centre for Nanofabrication (MCN) in the Victorian Node of the Australian National Fabrication Facility (ANFF).
D.E.G. acknowledges the ARC for support through a Future Fellowship (FT140100514). 
A.R acknowledges the ARC for support through a Discovery Project (DP110100221)

\section{Author contributions}
All authors contributed to writing the manuscript.
P.Z and T.S. performed the ultra--fast spectroscopy.
J.C. and A.R. measured the steady--state spectra.
D.C. and J.S. carried out simulations to identify the mechanical oscillations of the nanowires.
D.E.G. designed the experiments, fabricated the samples. 
D.E.G and T.D.  performed the theoretical analysis. 

\section{Competing financial interests}
The authors declare no competing financial interests

%
%
\cleardoublepage
\clearpage
\appendix

\renewcommand{\thesection}{S\arabic{section}}
\renewcommand{\thesubsection}{S\arabic{section}.\arabic{subsection}}
\renewcommand{\thefigure}{S\arabic{figure}}
\renewcommand{\theequation}{S\arabic{equation}}
\renewcommand{\thetable}{S\arabic{table}}
\setcounter{equation}{0}
\setcounter{figure}{0}

\section{Rigorous coupled wave analysis}
\label{sec:RCWA}

\begin{figure}[tbph!]
\centering
\includegraphics[width=0.9\linewidth]{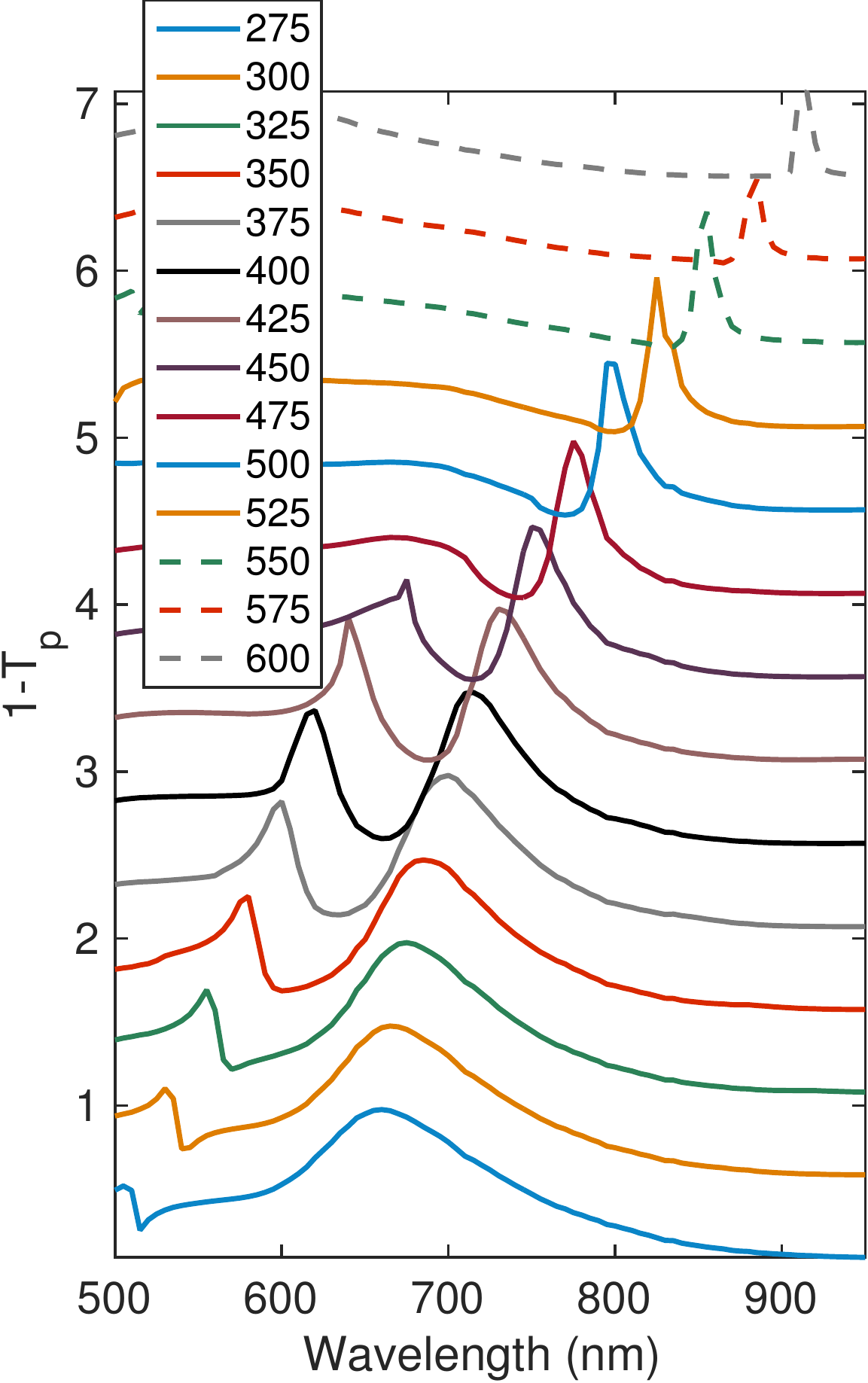}
\caption{Calculated 1 - T spectra for idealised Au nanowire gratings of width 100 nm and height 30 nm resting on top of 200 nm of TiO$2$.}
\label{fig:SExtinction_p_pol}
\end{figure}

\begin{figure}[tbph!]
\centering
(A)\\
\includegraphics[width=0.9\linewidth]{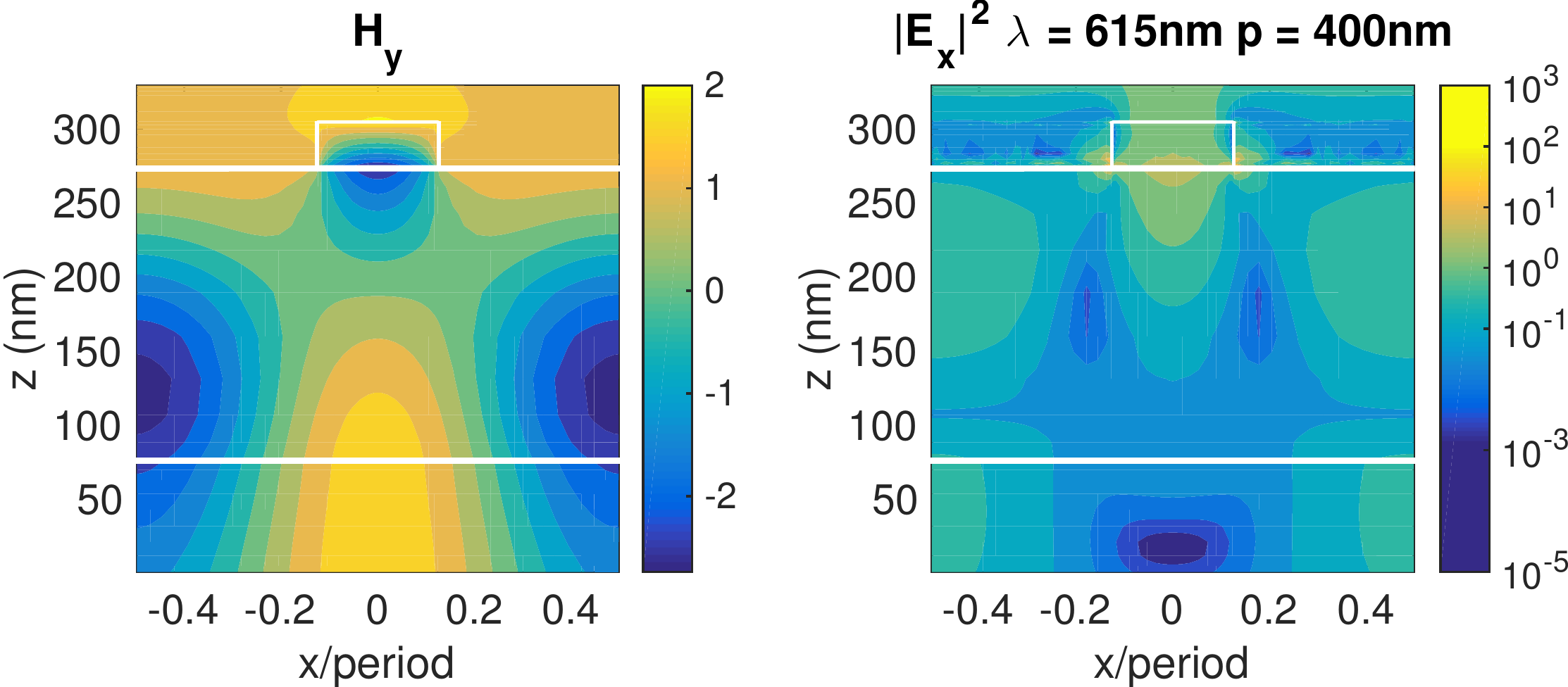}\\
(B)\\
\includegraphics[width=0.9\linewidth]{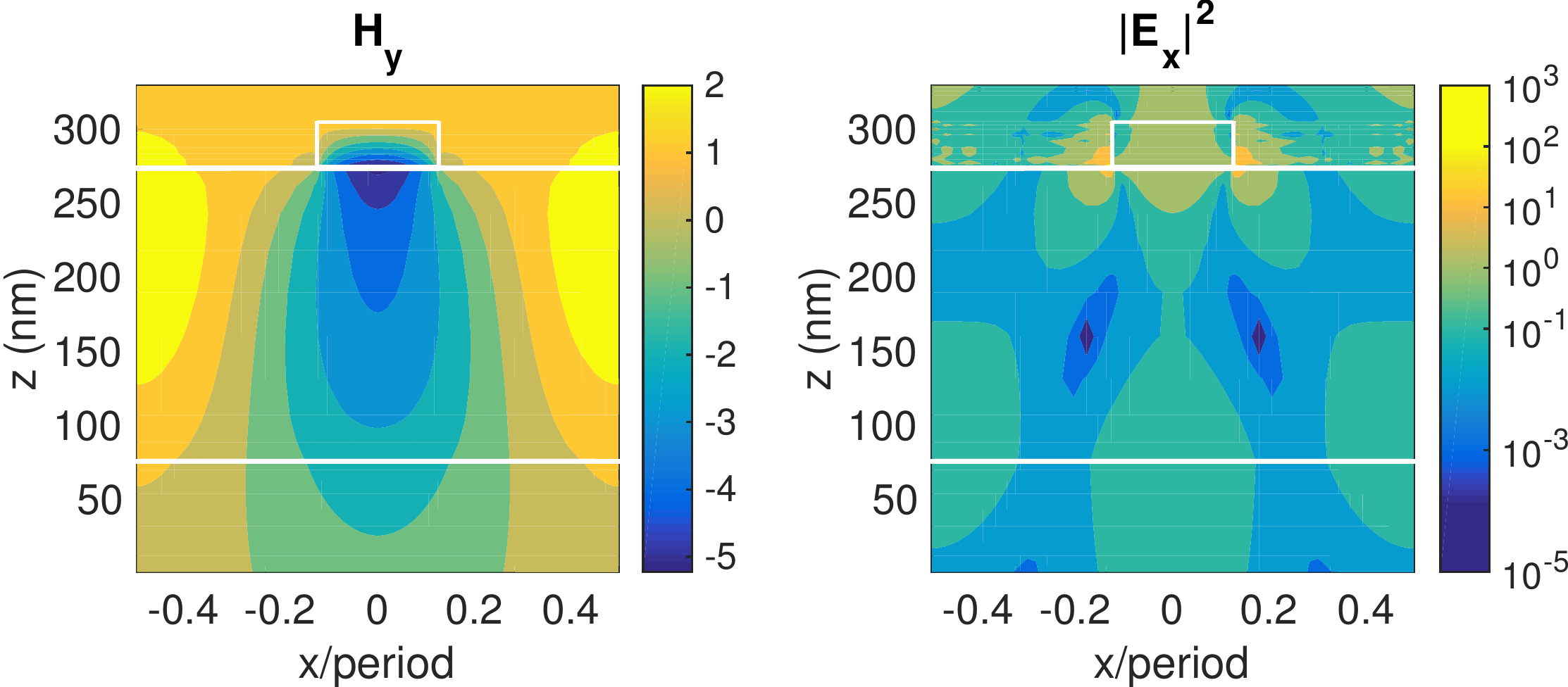}
\caption{Spatial map of the magnetic (left) and electric (field) on an Au nanowire supported on a 200 nm thick TiO$_2$ film for an incident wavelength of (A) 615 nm and (B) 710 nm and a nanowire grating period of 400 nm.}
\label{fig:s2}
\end{figure}

We employed the RETICOLO implementation of the frequency--domain Rigorous Coupled Wave Analysis for simulating the interaction of light with the  nanowire gratings \cite{Lalanne_JOMO1998a,Lalanne_JOSAA1996a,Moharam_JOSAA1995a}.
Convergence checks were carried out and allowed us to choose 50 Fourier orders for all the simulations shown here.
In the simulations we used tabulated data for the refractive index of Au\cite{Rakic_AO1998a}, assumed a refractive index of 1.5 for the substrate and 1. for the medium above the gratings 
The refractive index of the TiO$_2$ film was obtained from spectroscopic ellipsometry measurements.
Figure \ref{fig:SExtinction_p_pol} shows the calculated extinction spectra for $p$--polarisation, which exhibits good qualitative agreement with the experimentally measured data shown in figure 2(A).
For a grating period of 400 nm, we show in figure \ref{fig:s2} the maps of the magnetic and electric fields, the latter exhibitng a strong spatial confinement to the edges of the nanowires closest to the supporting TiO$_2$ films, in addition to a characteristic dipolar character. 
The map of the magnetic fields show also evidence of clamping of this fields at the position of the nanowires.
Both of these features are characteristic of waveguide--plasmon polaritons.

\begin{figure}[tbph!]
\centering
\includegraphics[width=0.9\linewidth]{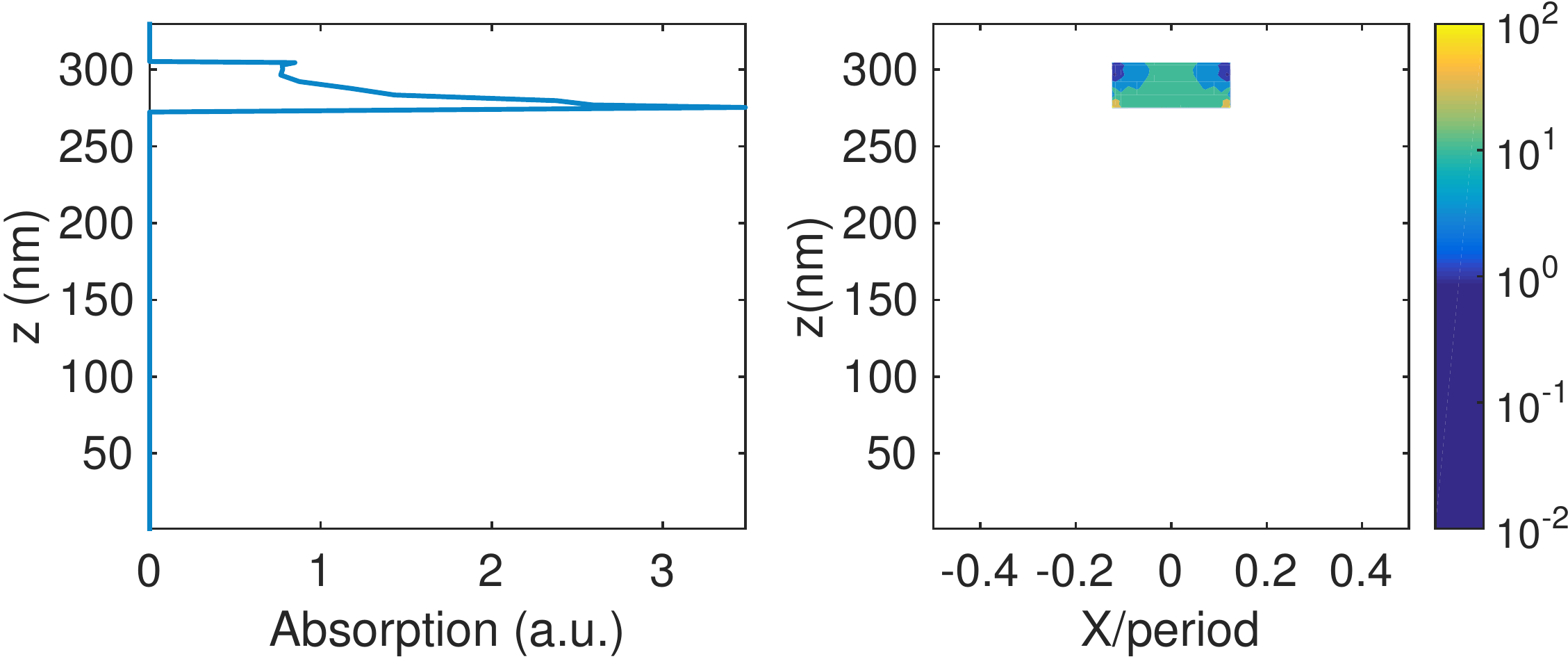}
\caption{(Left) Absorption loss {\it vs.} height ($z$) (Right) Map of the absorption loss.}
\label{fig:s3}
\end{figure}

In figure \ref{fig:s3} we show the ``absorption'' loss $L$, both as a function of height (i.e integarted over all other dimensions) and as a cross-section,  which is calculated for a wavelength $\lambda$ and for TM polarisation as:
\begin{equation}
L = \frac{\pi}{\lambda}\oint \text{Im}\{\epsilon(\lambda)\}\left(|E_x|^2 + |E_z|^2\right)dS,
\label{eq:S_L}
\end{equation}
where the integral is performed over a surface of a medium with a permittivity $\epsilon$.
This value is proportional to the amount of energy absorbed in the structure.
\begin{figure}[tbph!]
\centering
\includegraphics[width=0.9\linewidth]{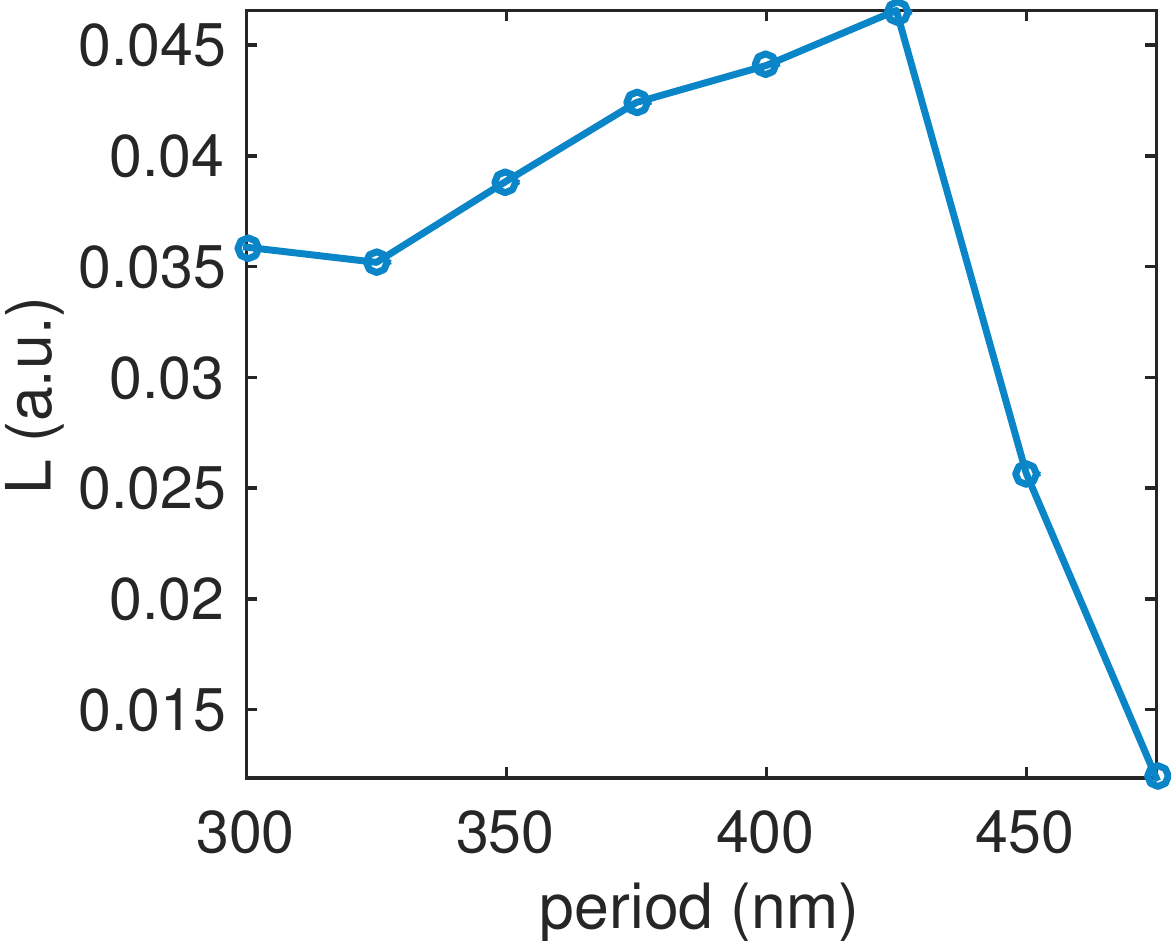}
\caption{$L$ of eqn. \eqref{eq:S_L} calculated for the higher--energy bands of figure \ref{fig:SExtinction_p_pol} shown as a function of grating period. }
\label{fig:L_vs_period}
\end{figure}

\section{Coupled oscillator model}
\label{sec:S_oscillators}
Here we present analytical expressions of the eigenvalues and eigenvectors of the Hamiltonian shown in equation \ref{eq:1}.

The uncoupled subsystems are idealised to share a quantum of excitation and consist of a  waveguide state denoted $|w\rangle$ and a plasmon state $|sp\rangle$.
Using these states as a basis set, the Hamiltonian reads
\begin{equation}
\begin{pmatrix}
E_p & \hbar\Omega/2 \\
\hbar\Omega^*/2 & E_W
\end{pmatrix}
-i\hbar
\begin{pmatrix}
\gamma_p & 0\\
0&\gamma_w
\end{pmatrix},
\label{eq:s1}
\end{equation}
where the second term accounts for damping in either sub-system.
Note that we are not considering cross-damping terms which have been shown to account for incoherent exchange of energy between the uncoupled sub-systems\cite{Wang_AN2014a}. 
The energy dispersion curve shown in figure 2(A) is given by the real part of the   eigenvalues of the Hamiltonian,
\begin{equation}
E_\pm(\delta) = \frac{\bar{E}_p+\bar{E}_W}2\pm \sqrt{\left(\frac{\bar{E}_p-\bar{E}_w}{2}\right)^2+|\hbar\Omega/2|^2},
\end{equation}
where $\bar{E}_k = E_k -i\hbar\gamma_k$.
The imaginary part of these eigenvalues gives in turn the expected linewidths of the polariton branches,a result we show in figure 2(C)

To find the eigenvalues, we shift, for the sake of simplicity,  all energy values with respect to $(E_p+E_W)/2$ defining a  $\delta = E_p-E_W$, rendering the real (Hermitian) part of the Hamiltonian:
\begin{equation}
\frac{1}{2}
\begin{pmatrix}
\delta & \hbar\Omega \\
\hbar\Omega^* & -\delta
\end{pmatrix}
\end{equation}
For which it is straightforward to show that the (normalised) upper polariton state $|P_+\rangle$ can be written as the following superposition of states:
\begin{equation}
|P_+\rangle = 
\frac{1}{\sqrt{1+\left(\frac{e_+ + 2\delta}{\hbar\Omega}\right)^2}}|w\rangle +
\sqrt{\frac{\delta + 2e_+}{2e_++\delta+\hbar\Omega}}|sp\rangle
\label{eq:S_cp}
\end{equation}
where :
\begin{equation}
e_+ = \sqrt{\delta^2+(\hbar\Omega)^2}/2
\end{equation}

\begin{figure}[tbph!]
\centering
\includegraphics[width=0.9\linewidth]{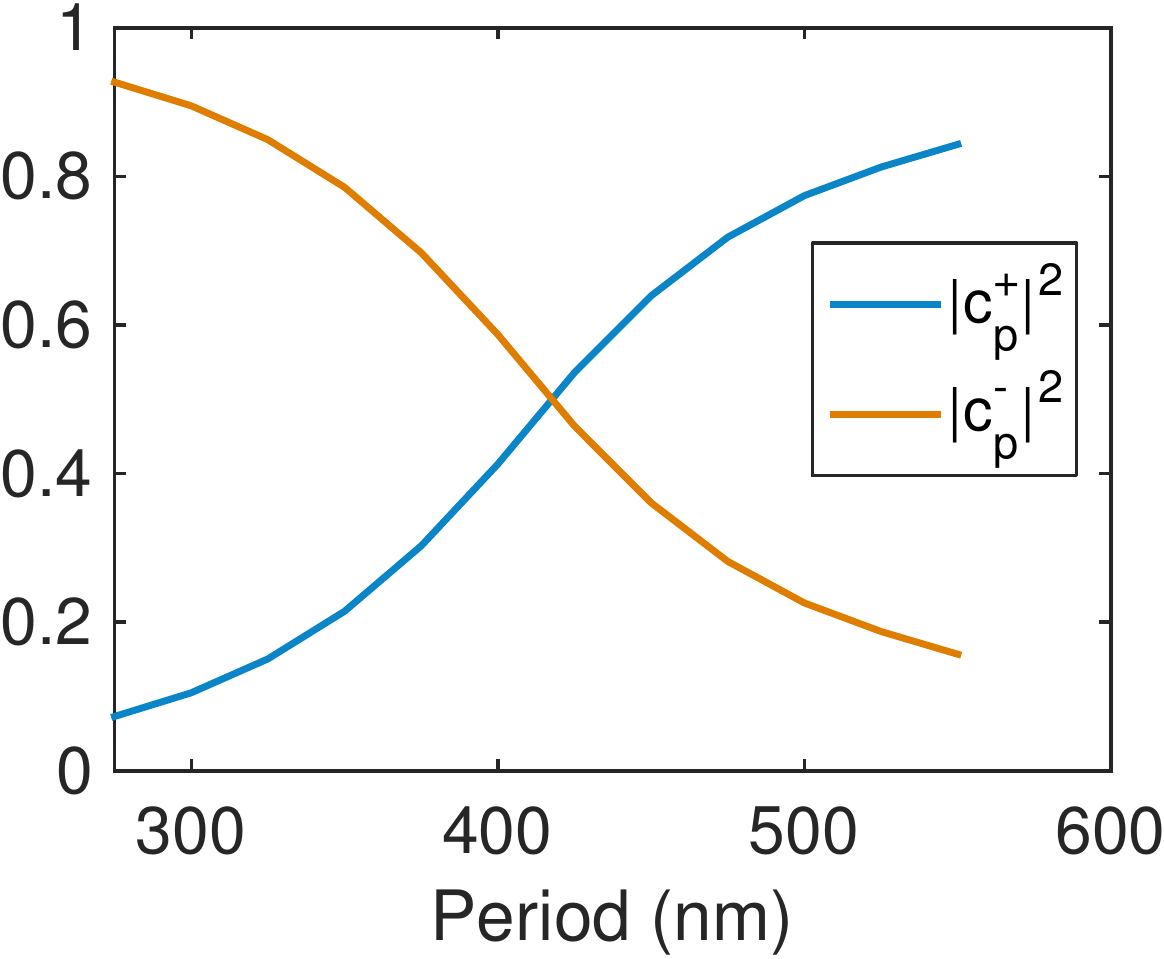}
\caption{Evolution of the plasmon coefficients {\it vs.} grating period obtained from the data shown in figure 2(B) using the model described in this section.}
\label{fig:SFig2}
\end{figure}

Figure \ref{fig:SFig2} shows the evolution of the ``plasmon coefficient'' (the factor before $|sp\rangle$) with grating period obtained for the data of figure 2.

\cleardoublepage
\section{Additional transient absorption results}
\label{sec:sTA}

\begin{figure}[tbph!]
\centering
\includegraphics[width=0.7\linewidth]{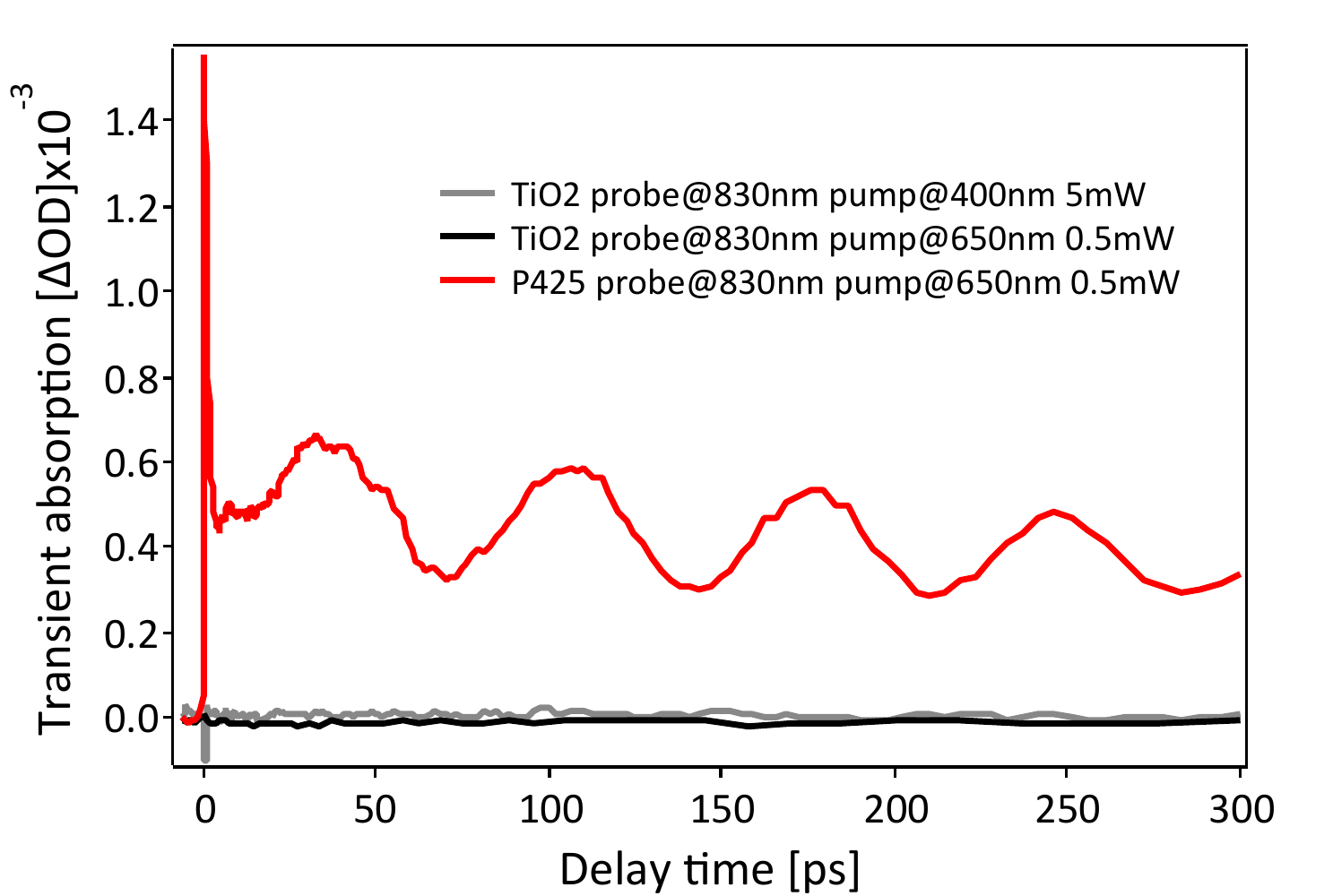}
\caption{Transient absorption measurements performed on a grating and its adjacent (bare) TiO$_2$ surfaces.}
\label{fig:S_reference}
\end{figure}

\begin{figure}[tbph!]
\centering
\includegraphics[width=0.7\linewidth]{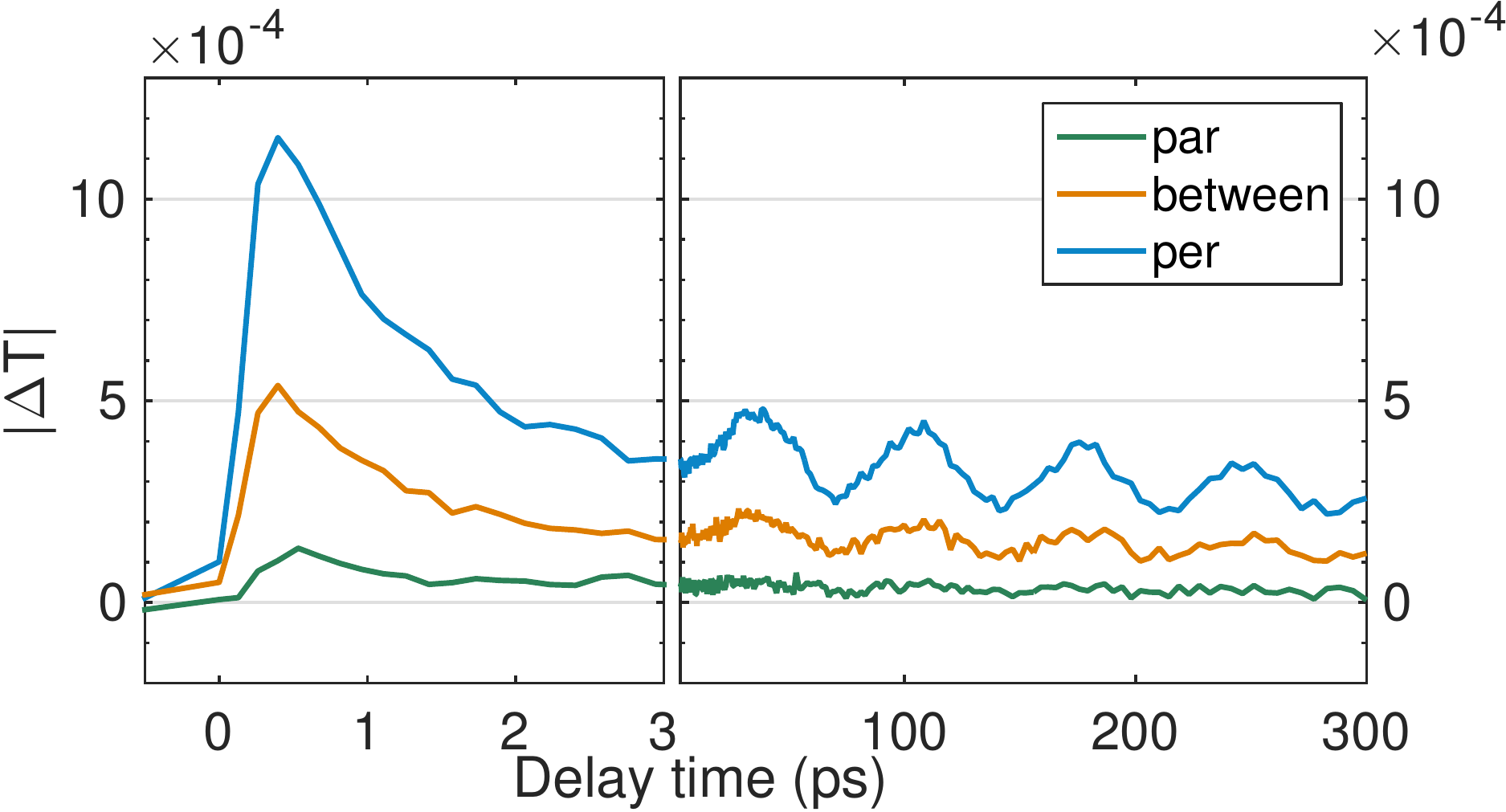}
\caption{Effect of the polarisation of the incident pump beam on the measured transient transmission spectrum. Grating period 425 nm.}
\label{fig:Spol}
\end{figure}

\begin{figure}[tbph!]
\centering
\includegraphics[width=0.7\linewidth]{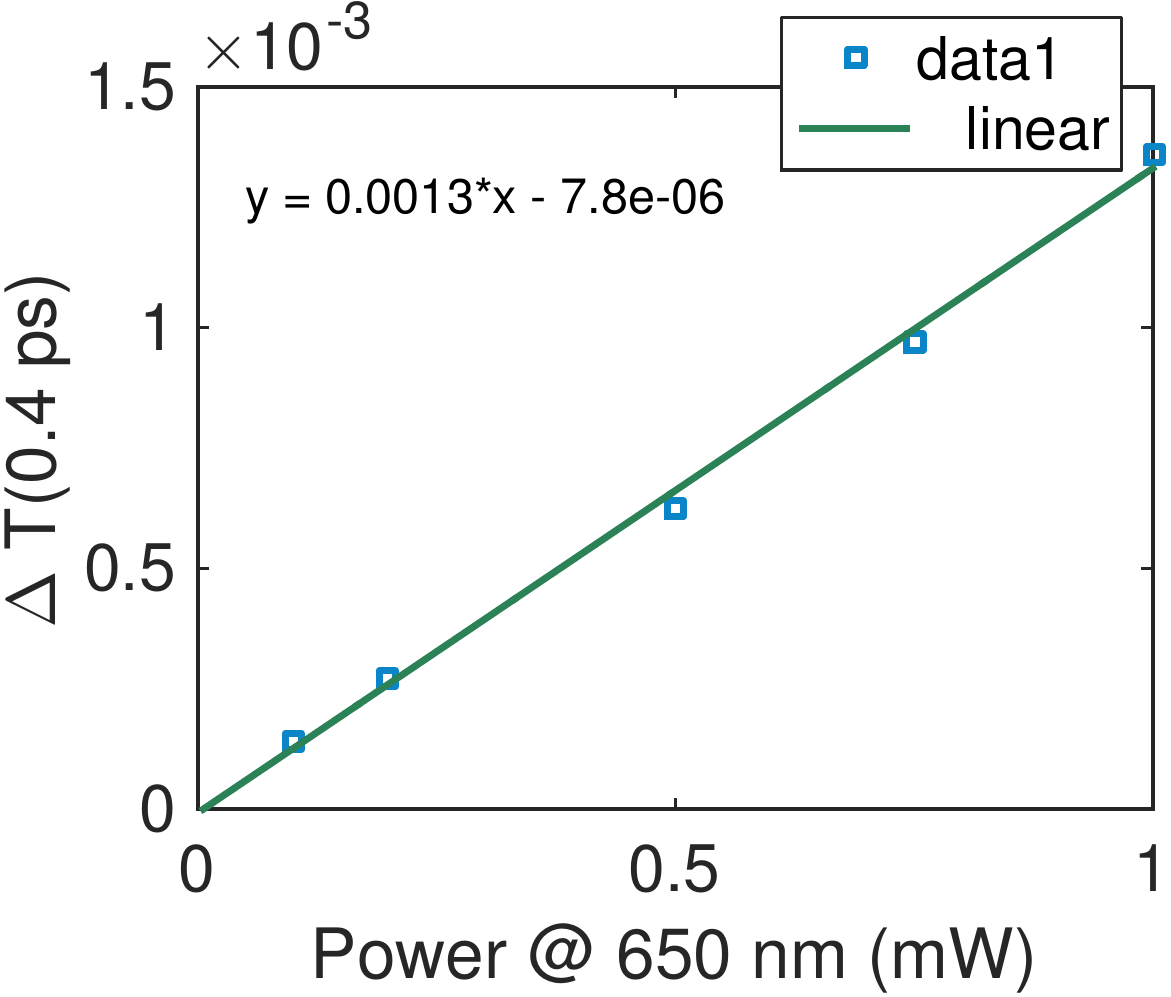}
\caption{Plot of the intensity of the transient signal vs pump power exhibiting a clear linear relationship. Grating period 425 nm.}
\label{fig:Spower}
\end{figure}

%
%

\begin{figure}[tbph!]
\centering
\includegraphics[width=0.7\linewidth]{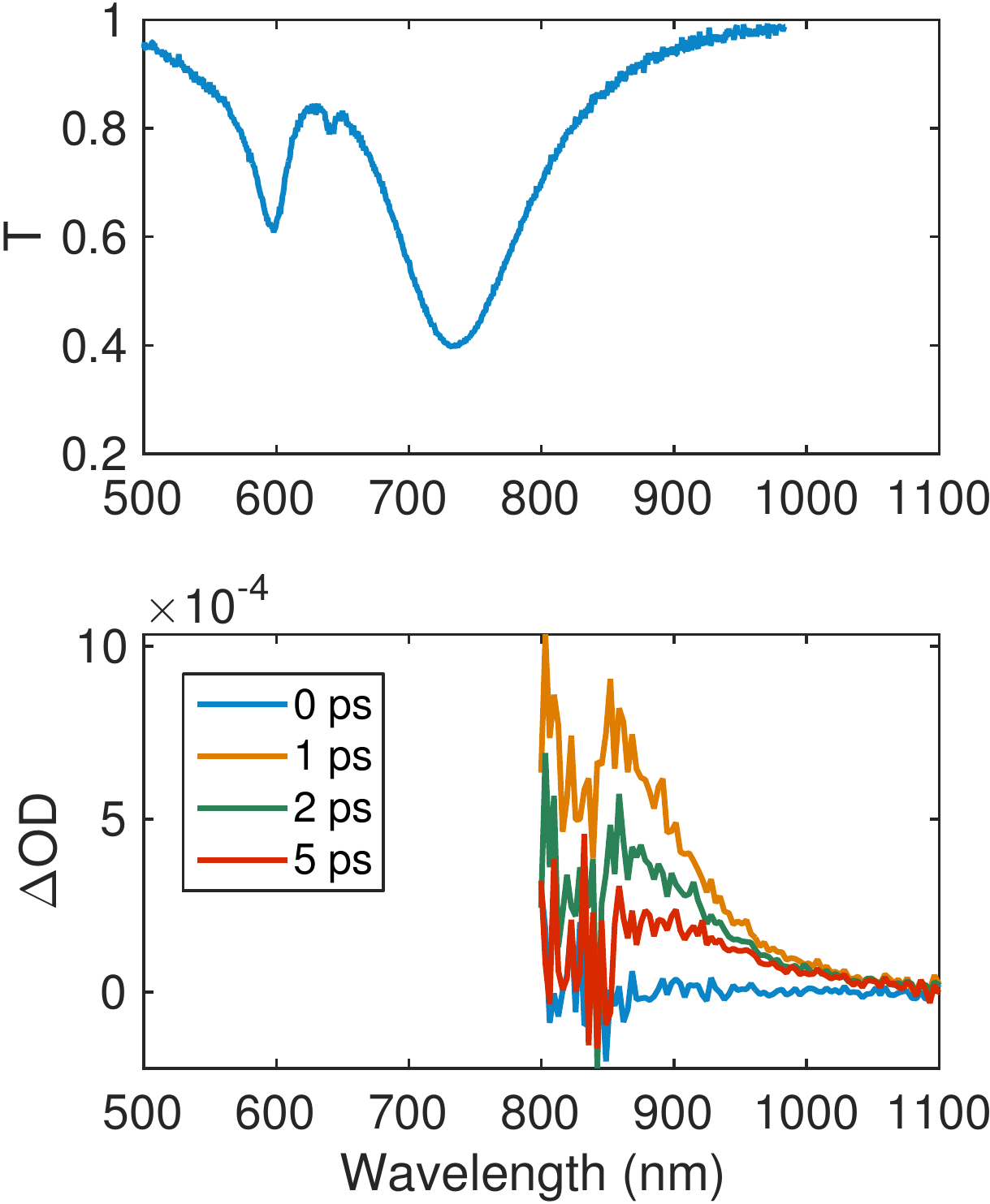}
\caption{(Top) steady--state and (Bottom) transient absorption spectra for a sample with a nanowire grating period of 350 nm.}
\label{fig:p350_Schottky_ss_ta_20151022}
\end{figure}
\begin{figure}[tbph!]
\centering
\includegraphics[width=0.7\linewidth]{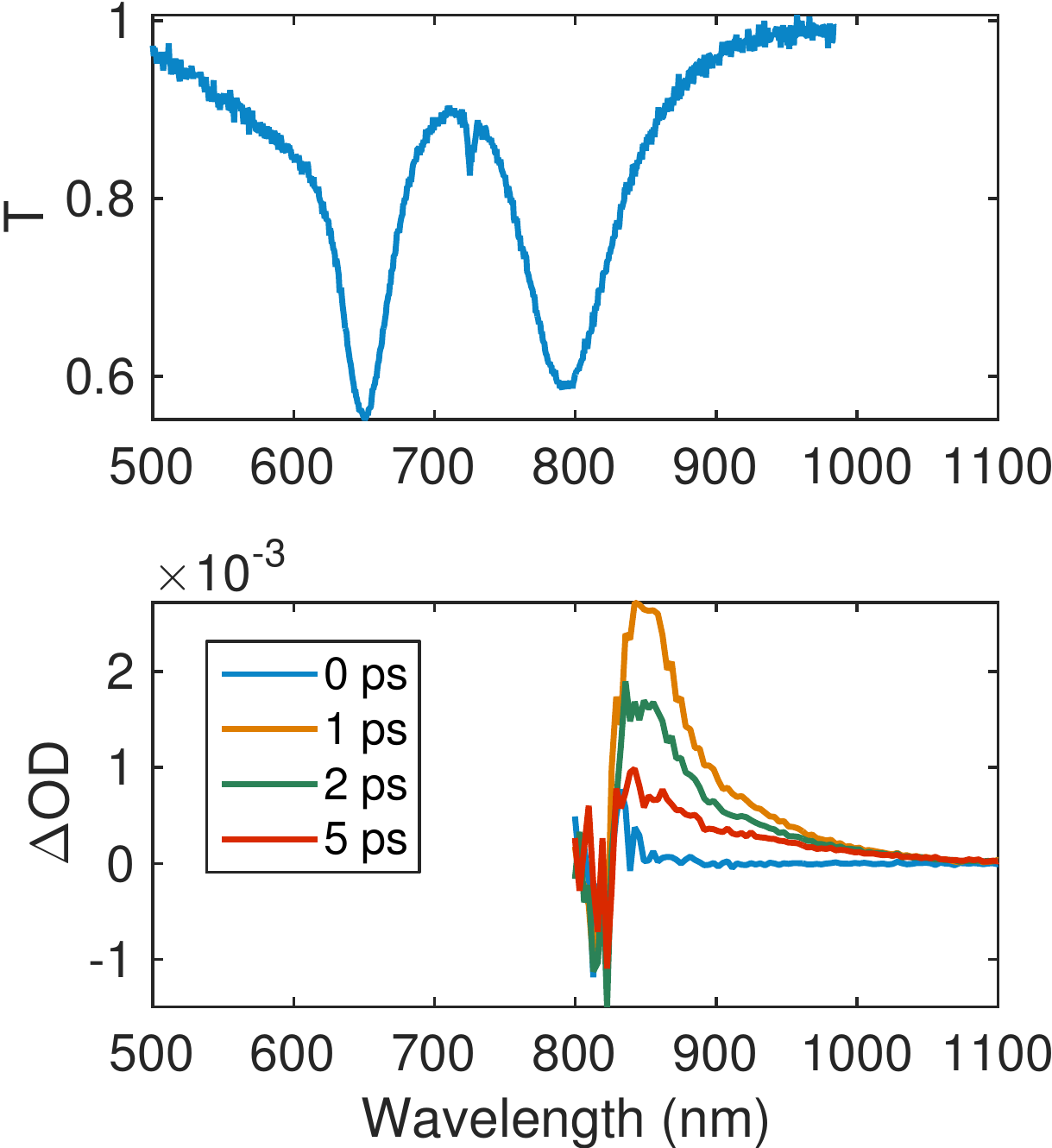}
\caption{(Top) steady--state and (Bottom) transient absorption spectra for a sample with a nanowire grating period of 425 nm.}
\label{fig:p425_Schottky_ss_ta_20151022}
\end{figure}
\begin{figure}[tbph!]
\centering
\includegraphics[width=0.7\linewidth]{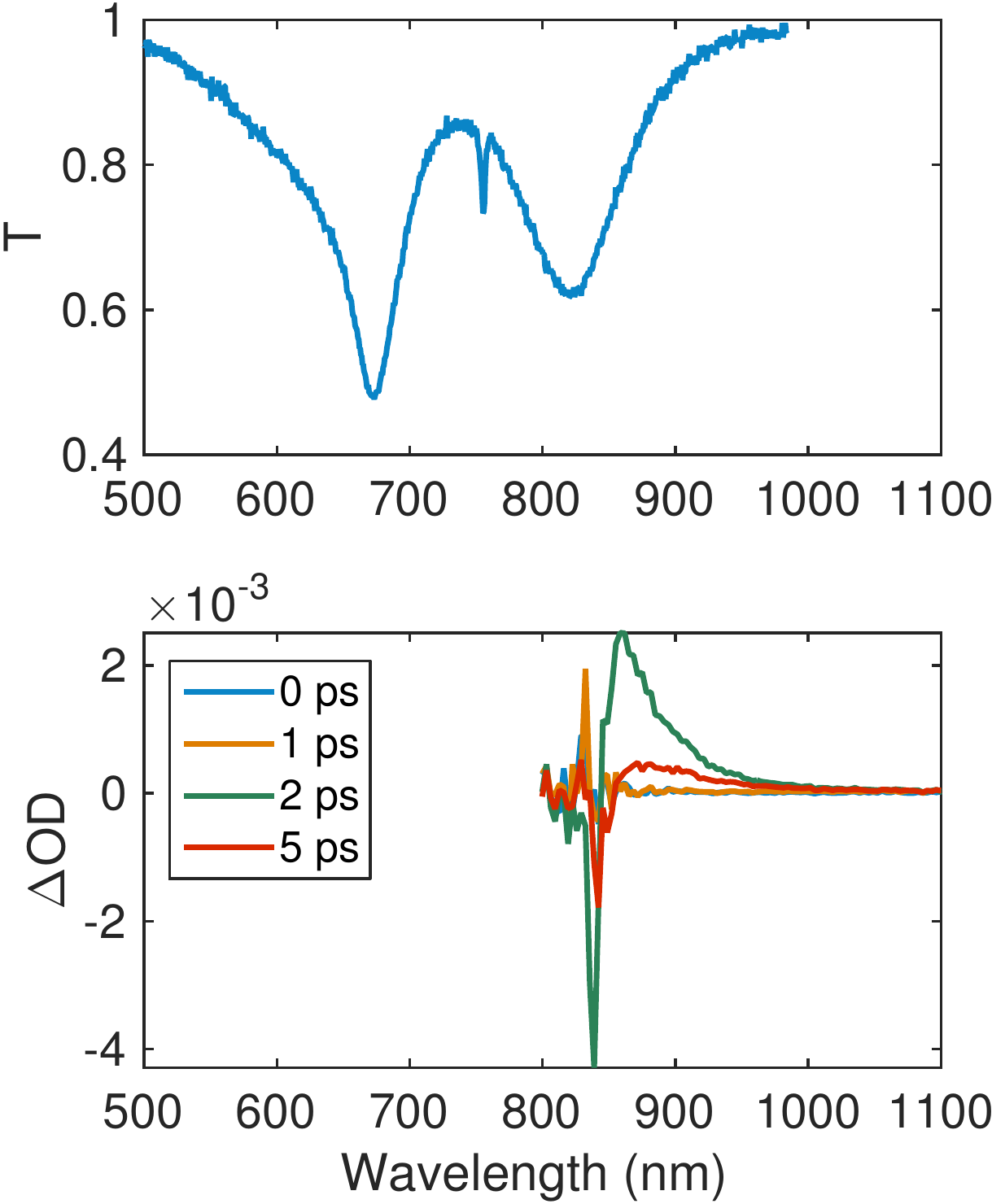}
\caption{(Top) steady--state and (Bottom) transient absorption spectra for a sample with a nanowire grating period of 450 nm.}
\label{fig:p450_Schottky_ss_ta_20150924}
\end{figure}
\begin{figure}[tbph!]
\centering
\includegraphics[width=0.7\linewidth]{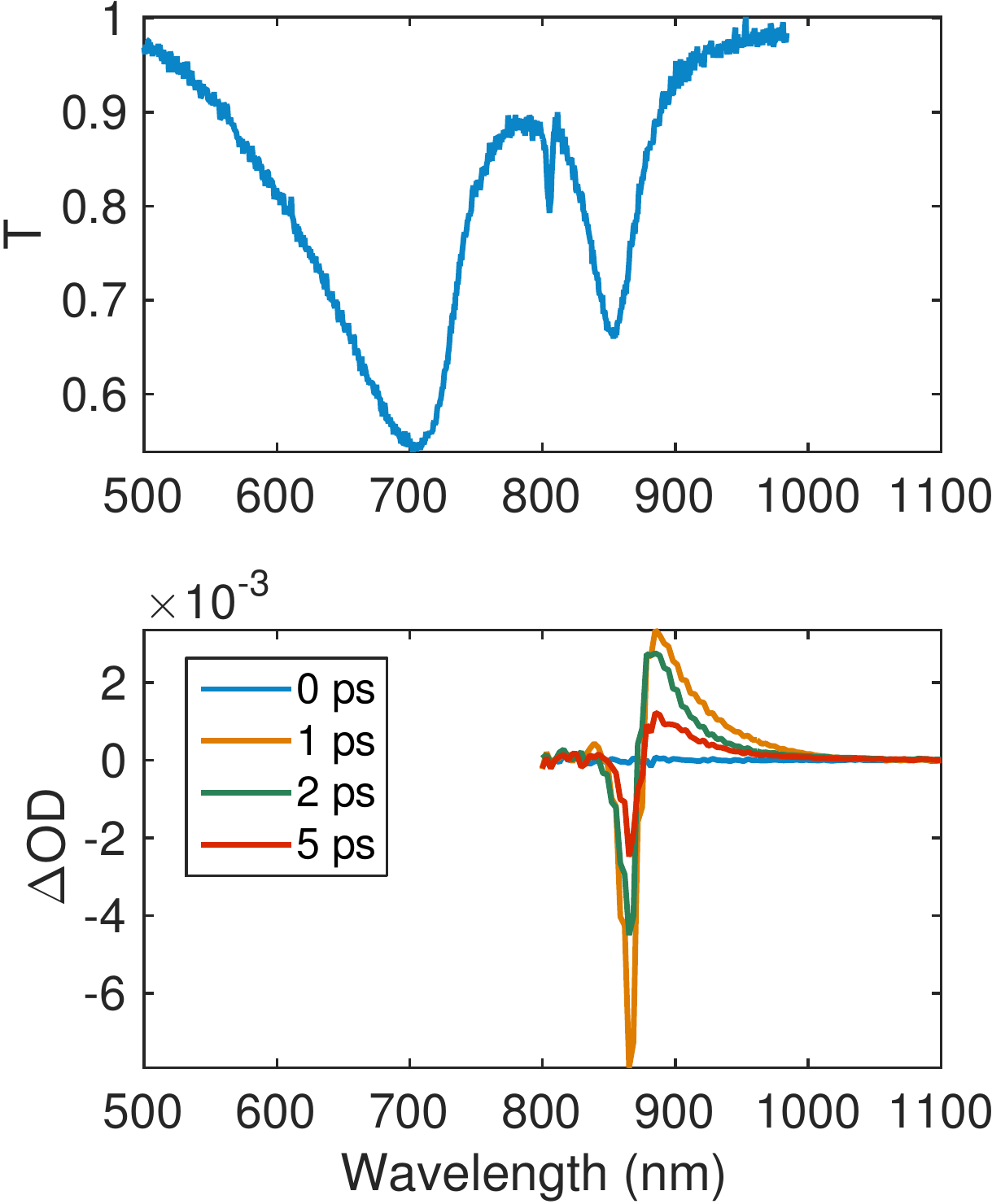}
\caption{(Top) steady--state and (Bottom) transient absorption spectra for a sample with a nanowire grating period of 500 nm.}
\label{fig:p500_Schottky_ss_ta_20150924}
\end{figure}
\begin{figure}[tbph!]
\centering
\includegraphics[width=0.7\linewidth]{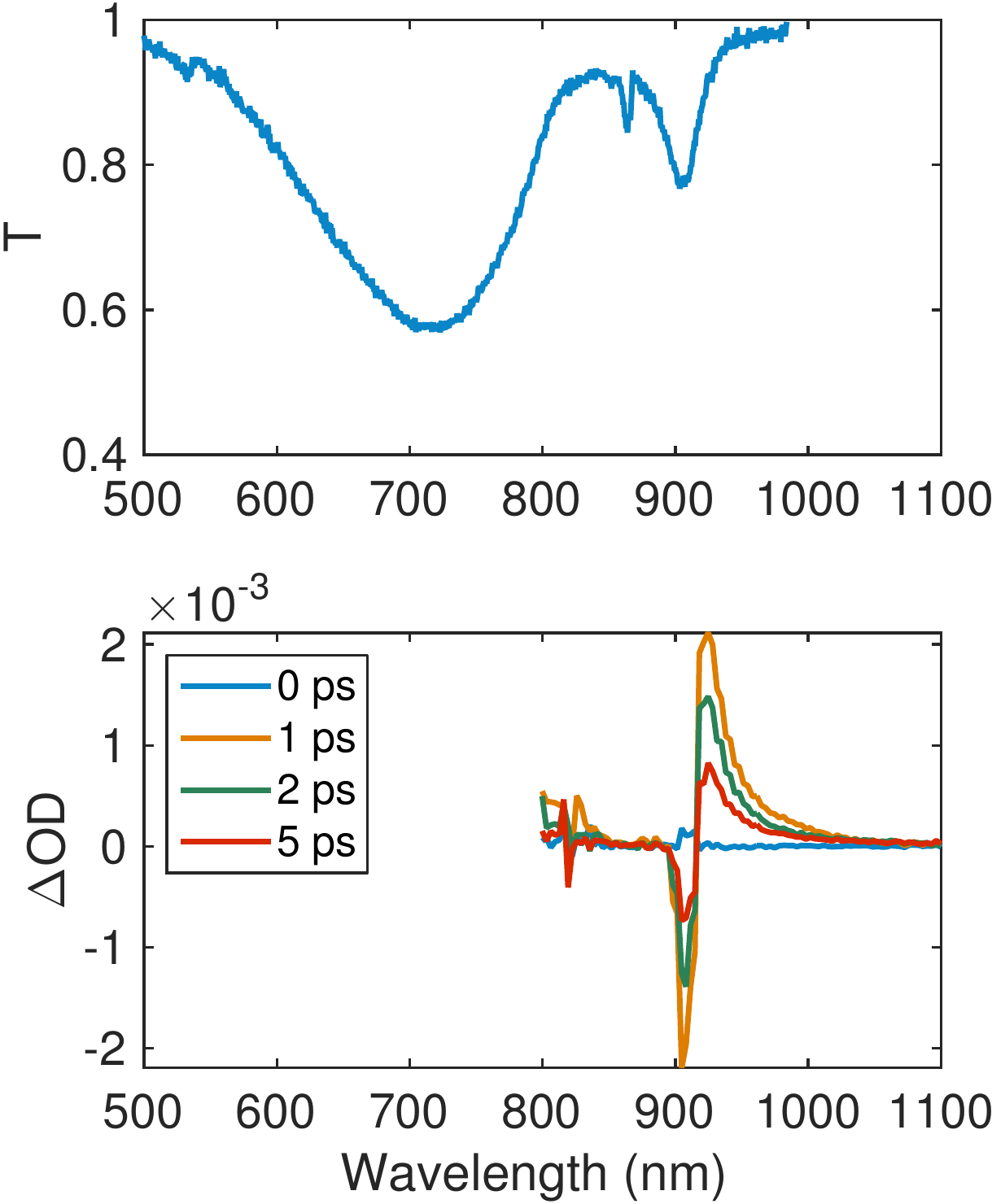}
\caption{(Top) steady--state and (Bottom) transient absorption spectra for a sample with a nanowire grating period of 550 nm.}
\label{fig:p550_Schottky_ss_ta_20150924}
\end{figure}

\cleardoublepage
\section{Assignment of mechanical oscillations}
\label{sec:SI_mech}

\begin{figure}[tbph!]
\centering
\includegraphics[width=0.9\linewidth]{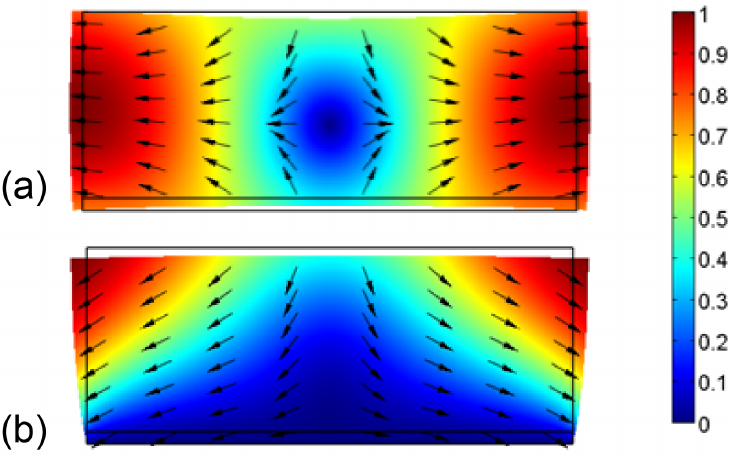}\\
\caption{FEM simulations of the acoustic vibration mode of single Au/Ti nanowire. Results of the displacement fields using a free surface model (a) and a fixed surface model (b). A cross-sectional view of the nanowire is presented, where arrows indicate the direction of the displacement field and thin black lines represent the unperturbed shape of the nanowire. The colour scale bar represents the normalized displacement field. (Bottom) Fourier transform of the data for the nanowire grating with a 425 nm period for t $>$ 10 ps, clearly displaying a peak at 14 GHz. This Fourier component can be accounted for  by coherent acoustic oscillations (transverse) of the Au nanowires.}
\label{fig:8}
\end{figure}

FEM simulations of the mechanical properties of the Au nanowires (including  Ti adhesion layers for Ohmic samples). 
We have considered breathing--mode oscillations of a cross section of the composite nanowire (see Fig \ref{fig:8})  under two limiting cases. 
In the first case, the bottom surface of the bimetallic nanowire is kept free, and in the second case the bottom surface of the bimetallic nanowire is rigidly fixed, i.e., no displacement is allowed. 
This covers the limiting cases where the substrate is not affecting nanowire motion, and when the substrate rigidly restrains it. 
FEM simulations predict breathing mode vibrational frequencies of 13.8 GHz and 14.7 GHz for weak and strong mechanical coupling with the  substrate, respectively. 
Experimental results indicate a vibrational frequency of 13.3 GHz which is in reasonable agreement with the two limiting cases we have considered.

\section{Kinetic model}
\label{sec:kinetic_model}

We consider an over--simplified situation where the initially populated $P_+$ polariton states decay through electron injection into accepting states in the semiconductor or by means of radiative or other non--radiative decay mechanisms.

After excitation of the system into a $P_+$ state, the population $N_A(t)$ of electrons in the accepting states of TiO$_2$ evolves according to the following differential equation:
\begin{equation}
\dot{N}_A = k_{inj} |c_p^+|^2 P_+ - k_{rec} N_A.
\end{equation}
The population of the polariton state $P_+$ instead decays according to:
\begin{equation}
\dot{P}_+ = -(k_{inj} + k)P_+
\end{equation}
where the dot represents a time--derivative and $k$ represents a sum over all rates causing a decrease in the population of the polariton state.

The population of electrons in the accepting states in the semiconductor $N_A$ increases by electron transfer from the polariton state $P_+$, in particular, from its plasmon--like component given by $|c_p^+|^2$, which according to the data shown in figure \ref{fig:2}, varies with the periodicity of the nanowire grating.
$N_A$ decreases in time due to back--electron transfer to the metal, a process that occurs with a rate constant $k_r$.

The population of the excited state decays via electron transfer to the accepting states with a rate given by $k_{inj}$ or by means of other radiative or non--radiative decay mechanisms. 
The equation describing the time evolution of $P_+$  can be readily integrated to yield an exponential decay:
\begin{equation}
P_+(t) = P_+(0) e^{-t(k_{inj} + k)},
\end{equation}
where $P_+(0)$ is proportional to the number of absorbed photons.

Integration of the  equation for $N_A(t)$, subject to $N_A(0)=0$, i.e. no electrons in accepting states at time $t=0$, yields:
\begin{equation}
N_A(t) = \phi_{inj} |c_p^+|^2 P_+(0)\left[e^{-k_rt} - e^{-(k_{inj} + k)t}\right],
\end{equation}
where the quantum yield of electron injection has been defined as:
\begin{equation}
\phi_{inj} =\frac{k_{inj}}{k_{inj} + k_{rad} + k_{nr}}.
\end{equation}

\begin{figure}[tbph!]
\centering
\includegraphics[width=0.9\linewidth]{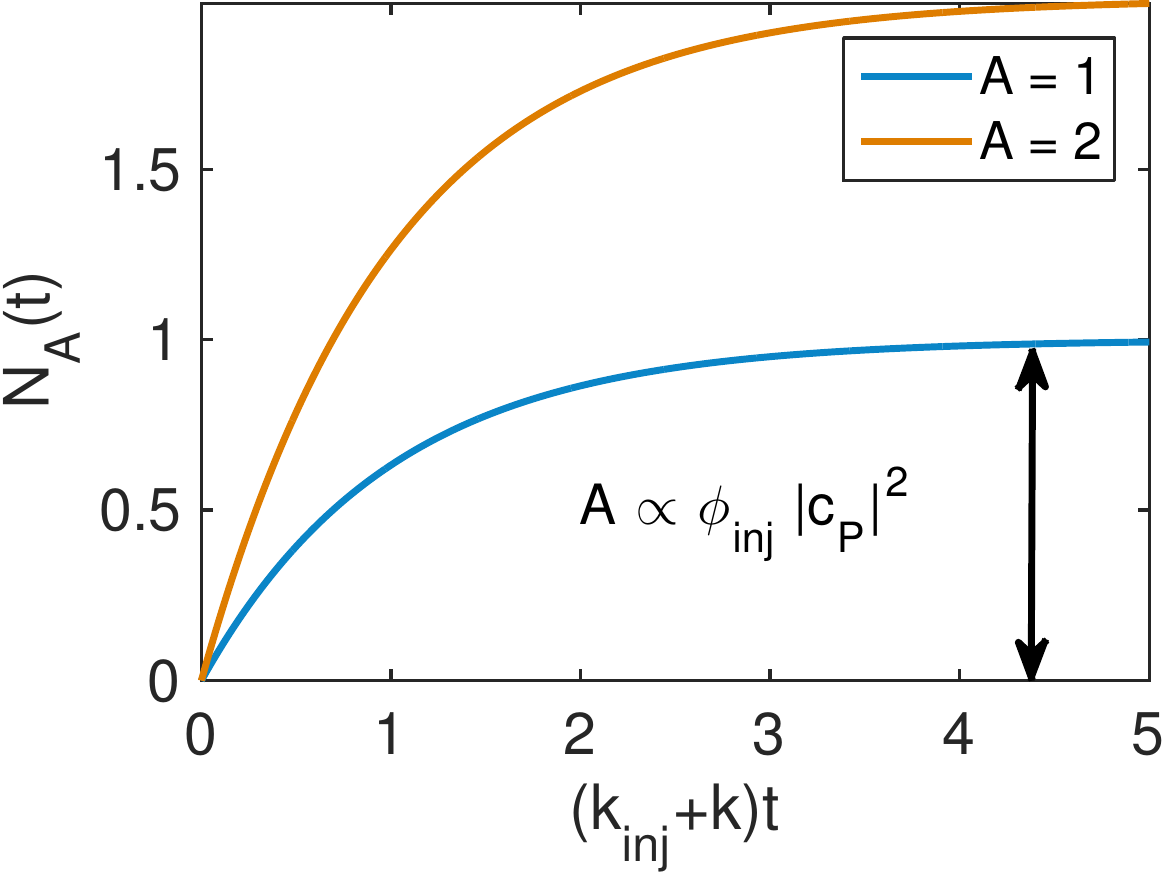}
\caption{Kinetic model.
Plot of $N_A$ {\it vs.} time (in units of $(k_{inj}+k)t$ for two values of the amplitude $A= \phi_{inj} |c_p^+|^2 P_+(0)$ as indicated. 
 }
\label{fig:kineric_model}
\end{figure}

For time scales much shorter than the recombination rate ($k_r\sim 1 \text{ns}^{-1} $, see Furube {\it et al} \cite{Furube_JOTACS2007a}) $N_A(t)$ is approximately given by:
\begin{equation}
N_A(t) \approx \phi_{inj} |c_p^+|^2 P_+(0)\left[1 - e^{-(k_{inj} + k)t}\right],
\label{eq:SNa}
\end{equation}
which implies that the maximum amplitude  of $N_A$ is proportional to $\phi_{inj} |c_p^+|^2 P_+(0)$, as indicated in figure \ref{fig:kineric_model}.
For experiments carried out under similar excitation densities (i.e. similar number of incident photons over the sample, and assuming similar absorption cross-sections), differences in the amplitudes of $N_A$ are proportional to $\phi_{inj} |c_p^+|^2$.
A similar model has been considered by Tian {\it et al}\cite{Tian_JOPAPAC2008a}.

\section{Radiative damping of a classical oscillator}
\label{sec:S_rad}

The rate of radiative damping in a classical oscillator is given by
\begin{equation}
k_\text{rad} = \frac{1}{4\pi\epsilon_b}\frac{4\omega^3|\mathbf{d}|^2}{3\hbar c^3}
\end{equation}
where $\mathbf{d}$ is the dipole moment of the oscillator, $c$ is the speed of light, $\omega$ the frequency and $\epsilon_B$ the permittivity of the medium surrounding the oscillator.

When this oscillator is coupled to another one, such as in the case of plasmon--waveguide strong coupling, the dipole moment $\mathbf{d}_\pm$ is given as a linear superposition of the dipole moments ($\mathbf{d}$ and $\mathbf{d}_c$) of the coupled oscillators\cite{Wang_AN2014a}:
\begin{equation}
\mathbf{d}_\pm = a_\pm \cdot \mathbf{d} + b_\pm \cdot \mathbf{d}_{c},
\end{equation}
where $a_\pm$ and $b_\pm$ are the mixing coefficients (similar to those encountered in \ref{sec:S_oscillators}), $\mathbf{d}_c$ is the effective dipole of the waveguide mode. 
If $\mathbf{d}_c$ is smaller than $\mathbf{d}$, then strong coupling results in a hybrid mode with a dipole moment that can be smaller than $\mathbf{d}$ leading to a reduction in the radiative decay rate.
\cleardoublepage

\end{document}